\documentclass[12pt]{iopart}

\usepackage{graphicx}% Include figure files
\usepackage{dcolumn}% Align table columns on decimal point
\usepackage{bm, amssymb, color}% bold math
\usepackage{hyperref}% add hypertext capabilities
\usepackage{booktabs,tabulary}
\usepackage{epstopdf}
\usepackage{soul}
\usepackage[caption=false]{subfig}
\usepackage{array}
\usepackage{cases}
\usepackage{multirow}

\usepackage[textwidth=16cm, top=1in, bottom=1in]{geometry}

\usepackage{enumitem}
\usepackage{amssymb} % A package for extra mathematical symbols.
\usepackage{bm} % A package for bold mathematical symbols.

\newcommand{\N}{\mathbb{N}} % Defines the command "\N" to be "N" written in blackboard bold font.
\newcommand{\Z}{\mathbb{Z}}
\newcommand{\R}{\mathbb{R}}

\newcommand{\C}{\mathbb{C}}

\newcommand{\fn}[2]{\mathinner{#1\mathopen{\left(#2\right)}}} 
\newcommand{\order}[1]{\mathinner{\mathcal{O}\mathopen{\left(#1\right)}}}
 
\newcommand{\abs}[1]{\left\vert #1 \right\vert}
\newcommand{\ceil}[1]{\lceil #1 \rceil}
\newcommand{\floor}[1]{\lfloor #1 \rfloor}

\newcommand{\vect}[1]{\bm{#1}}

\newcommand{\eqref}[1]{(\ref{#1})}
\newcommand{\pder}[2]{\mathop{\frac{\partial #1}{\partial #2}}}

\newcommand{\col}[1]{\fn{\tilde{n}}{#1}}
\newcommand{\colT}[1]{\fn{\tilde{n}_T}{#1}}

\newcommand{\Real}[1]{\mathrm{Re}\left[#1\right]}
\newcommand{\Imag}[1]{\mathrm{Im}\left[#1\right]}
\newenvironment{cond}
	{\left\{
    \begin{array}{l l}
    }
    { 
    \end{array} 
    \right.
    }

\begin{document}

\title{Inversion Problems for Fourier Transforms of Particle Distributions}% Force line breaks with \\

\author{Jaeuk Kim$^1$, Ge Zhang$^2$, Frank H. Stillinger$^3$, and Salvatore Torquato$^{1,3,4,5}$}

\address{$^1$
Department of Physics, Princeton University, Princeton, New Jersey 08544, USA}
\address{$^2$
Department of Physics and Astronomy, University of Pennsylvania, Philadelphia, Pennsylvania  19104-6396, USA}
\address{$^3$
Department of Chemistry, Princeton University, Princeton, New Jersey 08544, USA}
%\homepage{http://chemlabs.princeton.edu/torquato}
\address{$^4$\textbf{•}
Princeton Institute for the Science and Technology of Materials, Princeton University, Princeton, New Jersey 08544, USA}
\address{$^5$
Program in Applied and Computational Mathematics, Princeton University, Princeton, New Jersey 08544, USA}
\ead{torquato@electron.princeton.edu}

\date{\today}% It is always \today, today,
\begin{abstract}
Collective coordinates in a many-particle system are complex Fourier components of the particle density $\fn{n}{\vect{x}} \equiv \sum_{j=1}^N \fn{\delta}{\vect{x}-\vect{r}_j}$, and often provide useful physical insights.
However, given collective coordinates, it is desirable to infer the particle coordinates via inverse transformations.
In principle, a sufficiently large set of collective coordinates are equivalent to particle coordinates, but the nonlinear relation between collective and particle coordinates makes the inversion procedure highly nontrivial.
Given a ``target'' configuration in one-dimensional Euclidean space, we investigate the minimal set of its collective coordinates that can be uniquely inverted into particle coordinates.
For this purpose, we treat a finite number $M$ of the real and/or the imaginary parts of collective coordinates of the target configuration as constraints, and then reconstruct ``solution'' configurations whose collective coordinates satisfy these constraints.
Both theoretical and numerical investigations reveal that the number of numerically distinct solutions depends sensitively on the chosen collective-coordinate constraints and target configurations.
From detailed analysis, we conclude that collective coordinates at the $\ceil{\frac{N}{2}}$ smallest wavevectors is the minimal set of constraints for unique inversion, where $\ceil{\cdot}$ represents the ceiling function.
This result provides useful groundwork to the inverse transform of collective coordinates in higher-dimensional systems.

\end{abstract}

\pacs{xxx,xxx}
\vspace{2pc}
\submitto{\JSTAT}             
             
\maketitle

\section{Introduction}\label{sec:intro}
For $N$ identical point particles at positions of $\vect{r}_1, \cdots , \vect{r}_N$ in a periodic fundamental cell $\Omega$, the particle distribution can be described by a particle density $\fn{n}{\vect{x}} \equiv \sum_{j=1}^N \fn{\delta}{\vect{x}-\vect{r}_j}$.
Equivalently, this function can be represented by the (complex) Fourier components at wavevectors $\vect{k}$'s, associated with the geometry of $\Omega$, i.e.,
\begin{equation}\label{def:collective coordinates}
\col{\vect{k}} \equiv \sum_{j=1}^N e^{-i\vect{k}\cdot \vect{r}_j},
\end{equation}
called \textit{collective coordinates}.
These quantities are often found to be a natural way to describe the distribution of particles, and thereby provide useful insights into many physical problems, e.g., excited states of liquid helium \cite{Feynman1954}, conduction electrons in metals \cite{Pines1952}, general theory of simple liquids \cite{Percus1958}, and quantification of density fluctuations \cite{Torquato2003_hyper, Torquato2015_stealthy}.
Furthermore, using functional Fourier transformation, governing equations of many-body systems, such as the Fokker-Planck equation, can be expressed in terms of collective coordinates \cite{Edwards2003}.

It is often desirable to infer the particle coordinates from given collective coordinates via inverse transformations.
Importantly, amplitudes of collective coordinates, or equivalently, \textit{structure factors} $\fn{S}{\vect{k}}$'s have long been used to probe the particle distributions since $\fn{S}{\vect{k}}$ can be measured by scattering experiments \cite{Ashcroft}.
However, unless the particle distribution is a perfect crystal, the structure factor alone cannot uniquely determine the particle distribution.
To solve this problem in X-ray crystallography, additional information is acquired from other physical properties, such as the interference pattern with known molecules (specific site labeling) \cite{Kam1977}, anomalous dispersion relations \cite{Pahler1990, Hendrickson1997}, or sequential projections onto constrained hyperplanes \cite{Elser2003}.
Such inversion tasks are called the \textit{phase-retrieval problems} \cite{Elser2003,Shechtman2015,  Harrison1993} because the tasks are essentially equivalent to retrieving the ``phase'' information contained in collective coordinates, the complete set of which are in principle invertible into particle coordinates.
Even if the phase information is incorporated, however, this inversion task is still highly nontrivial, due to the nonlinear relation between collective and particle coordinates.

Given a \textit{target} point configuration in one-dimensional Euclidean space $\R$, our primary objective in this paper is to find the minimal set of its collective coordinates that uniquely determine particle coordinates under exchange of particle indices.
This minimal set, therefore, uniquely determines collective coordinates at other wavevectors.
To carry out this search, we treat the number $M$ of the real and/or the imaginary parts of collective coordinates of a target configuration as \textit{constraints}, and find all configurations, called \textit{solutions}, whose collective coordinates satisfy these constraints.
The number of constraints $M$ is increased one-by-one until we have a unique solution that is, of course, identical to the target pattern.

\begin{figure}[ht]
\center
\includegraphics[width = 0.6\textwidth]{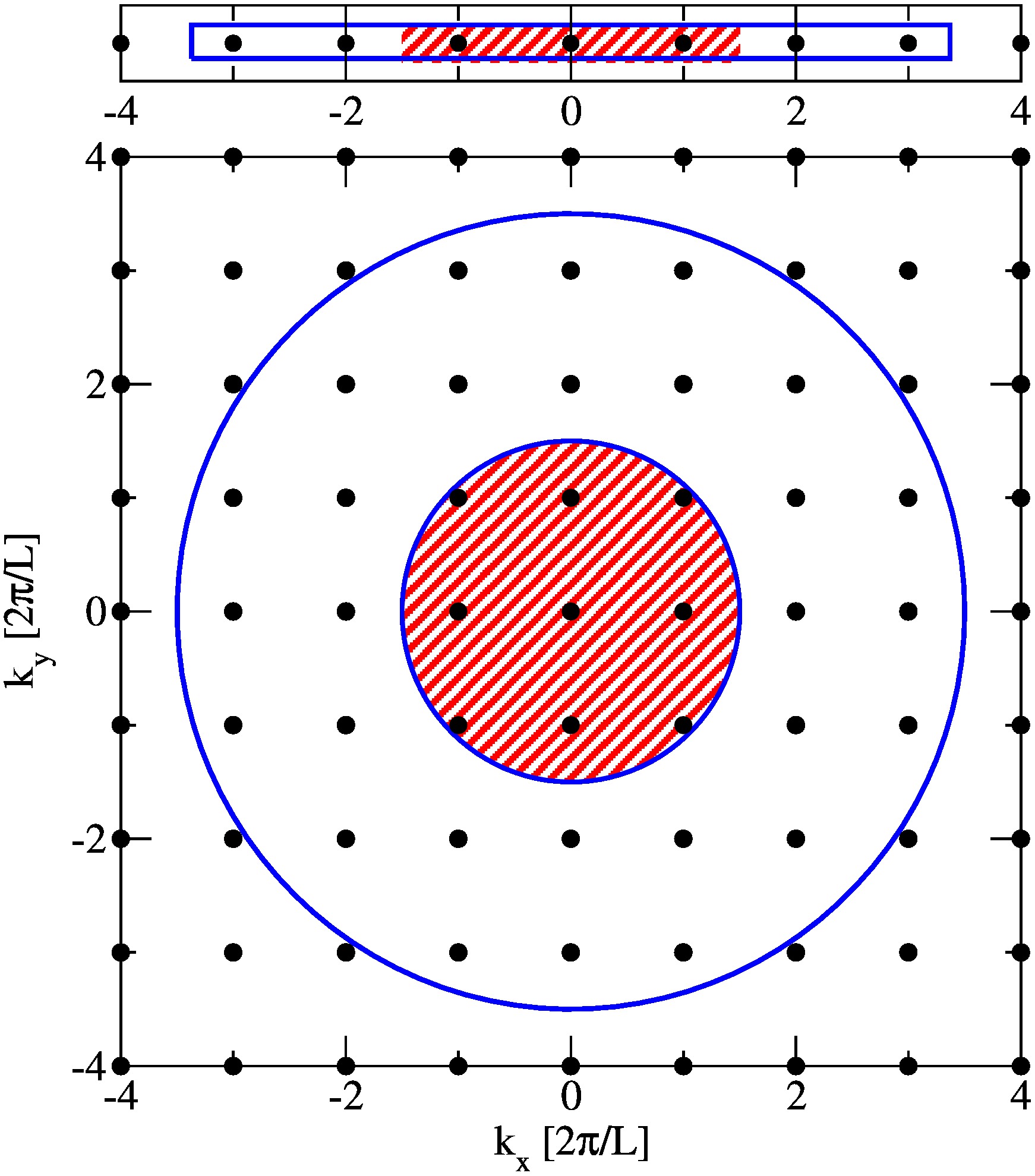}
\caption{Schematics of typical arrangements of collective-coordinate constraints in Fourier space for a periodic $d$-dimensional square fundamental cell of side length $L$. Here, upper and lower panels represent cases for $d=1$ and $2$, respectively.
Constraints are taken from $\col{\vect{k}}$'s at wavevectors between two concentric circles centered at the origin: there are $2\mathcal{M}$ wavevectors (black dots) within the blue circle, except for $2N_k +1$ wavevectors inside the red-shaded region.
%Then, the corresponding relative fraction of constrained degrees of freedom is denoted by $\chi = \mathcal{M}/[d(N-1)]$.
In Refs. \cite{Fan1991, Uche2004, Torquato2015_stealthy,Zhang2015, Zhang2015b}, a spherical region with $N_k=0$ was considered; see a list of available $\mathcal{M}$ values for two-dimensional cases in Table II in Ref. \cite{Uche2004}.
For our present purposes, the number of constraints is denoted by $M=2\mathcal{M}$ because the real and/or the imaginary parts of collective coordinates are considered independently.
\label{fig:Constraints}}
\end{figure}

Previous studies on this inversion task \cite{Torquato2015_stealthy, Fan1991, Zhang2015, Zhang2015b} focused on some special types of constraints in collective coordinates for a given set of wavevectors, such as the \textit{stealthy} constraints, where $\col{\vect{k}}=0$, and amplitude-constraints for a prescribed radial function $\fn{f}{r}$, i.e., $\abs{\col{\vect{k}}} = \fn{f}{\abs{\vect{k}}}$.
This inversion task is often carried out via the \textit{collective-coordinate optimization technique} \cite{Zhang2015, Zhang2015b,Uche2004, Uche2006,  Zhang2017c} that is designed to find ground-state configurations of the potential associated with those constraints.
Here, it is useful to define  a new parameter $\chi \equiv \mathcal{M}/(dN)$ \cite{Zhang2015, Uche2004} that represents the relative fraction of the number of constrained collective coordinates $\mathcal{M}$ to the total number of degrees of freedom; see figure \ref{fig:Constraints} for typical arrangements of the constraints in $d=1,2$.
These studies analytically or numerically showed that when the stealthy constraints are imposed for $\chi <1/2$, the associated ground states, called stealthy disordered hyperuniform systems \cite{Torquato2015_stealthy, Zhang2015,Zhang2015b,Uche2004}, are  disordered, highly degenerate, and statistically isotropic.
Importantly, it has been shown that systems, derived from these special disordered point configurations by decorating the points with particles of certain shapes, are endowed with some novel photonic and transport properties \cite{Florescu2009, Zhang2016, Zhang2017, Chen2017, Leseur2016, Scheffold2017, Gkantzounis2017, Wu2017}; see also Ref. \cite{Torquato2018_review} and references therein.
Under the stealthy constraints with $\chi\geq 1/2$, on the other hand, (virtually all) available configurations are crystalline in the first three spatial dimensions \cite{Torquato2015_stealthy, Fan1991,Uche2004}.
From the uniqueness of the solution at $\chi=1/2$ in $d=1$ \cite{Fan1991} as well as the importance of phase information of collective coordinates, one can argue  that each constrained collective coordinate $\col{\vect{k}}$ removes \textit{two} degrees of freedom in the accessible configurational space.
Thus, it is natural to surmise that the minimum value of $M$ for the unique inversion would be $M=dN$.

In the present work, we consider more general type of constraints, in which  the real and/or the imaginary part of each collective coordinate are independently prescribed.
For simplicity, we focus on one-dimensional systems. For such systems, we show that the minimal set of collective-coordinate constraints consists of collective coordinates at the $\ceil{N/2}$ smallest wavevectors, i.e., $M=2\ceil{N/2}$ rather than $N$.
This result also implies that for a collective coordinate at a wavevector $\vect{k}$, both its real and imaginary parts must be specified.
We analytically show this result for small systems of $N\leq 3$.
However, this result is invalid if the target configurations are the integer lattice because one cannot determine its center of mass without a collective coordinate at the first Bragg peak.
In our numerical studies for larger systems, we exclude the pathological case (i.e., the integer lattice), and consider two distinct ensembles of target configurations: perturbed lattices \cite{Gabrielli2004} via uniformly distributed displacements, and Poisson point distribution configurations.
For each of these target configurations, we find solutions numerically via the collective-coordinate optimization technique.
Our numerical results show that these two types of ensembles occupy qualitatively different energy landscapes: those in perturbed lattices are relatively simpler than those in Poisson ones.

In section \ref{sec:background}, we present basic definitions and background.
In section \ref{sec:numerical methods}, we describe the numerical method that we employ to find solutions.
In section \ref{sec:small systems}, we theoretically and numerically determine the minimal sets of collective coordinates for small systems.
Larger systems are numerically investigated in section \ref{sec:N>3}.
Finally, we provide concluding remarks in section \ref{sec:conclusions}.

\section{Basic Definitions and Background}\label{sec:background}
\subsection{General Properties of Collective Coordinates}\label{sec:collective coordinates}

For a $N$-particle point configuration within a periodic fundamental cell $\Omega$% in $d$-dimensional Euclidean space $\R^d$
, collective coordinates \eqref{def:collective coordinates}, which are also known as \textit{collective density variables}, are complex-valued quantities that are defined at certain real-valued discrete wavevectors $\vect{k}$'s.
Here, the available wavevectors correspond to the reciprocal lattice vectors of the cell $\Omega$.
For instance, if $\Omega$ is a $L_1 \times \cdots\times L_d$ rectangular box, then $\vect{k}$'s can be described as follows: $\vect{k}=2\pi(\frac{m_1}{L_1},\cdots, \frac{m_d}{L_d})$ for $(m_1,\cdots, m_d)\in \Z^d$.
For the simplicity, we focus on one-dimensional systems in the rest of this paper, and thus use the following short-hand notation:
\begin{equation}\label{eq:k_m}
k_m  = 2\pi m/L.
\end{equation}

At two different wavevectors, the collective coordinates are not always independent.
For instance,
the complex conjugate of a collective coordinate by definition is equal to its parity inversion, i.e., $\fn{\tilde{n}^*}{\vect{k}}=\col{-\vect{k}}$.
Thus, if we constrain such a pair of collective coordinates, only one of them is considered independent.
For this reason, the relative fraction $\chi$ of constrained degrees of freedom is defined as not $2\mathcal{M}/(dN)$, but $\mathcal{M}/(dN)$; see figure \ref{fig:Constraints}.

Only certain sets of complex numbers can be collective coordinates of a ``realizable'' point configuration.
For example, there are some trivial necessary conditions of realizable collective coordinates, such as $\abs{\col{\vect{k}}} \leq N$ for any wavevector $\vect{k}$, and $\col{\vect{0}} = N$.
However, it is highly nontrivial to find sufficient and necessary conditions of realizable collective coordinates.
To avoid such realizability problems \cite{Kuna2007}, we take constraints from the collective coordinates of a target configuration.

The value of a collective coordinate is independent of the choice of particle permutations: When we invert collective coordinates, the resulting particle coordinates also should be equivalent under exchange of particle indices.

\subsection{Definitions}\label{sec:def}
In the rest of this work, we clearly distinguish a target and a solution configurations by using separate notations $\vect{R}^N = \{R_1,R_2, \cdots, R_N \}$ and $\vect{r}^N = \{r_1,r_2, \cdots, r_N \}$, respectively.
The corresponding collective coordinates are denoted by $\colT{k}$ and $\col{k}$, respectively.

In numerical studies, two types of target configurations at unit number density are considered: 
\begin{enumerate}
\item
Perturbed lattices \cite{Gabrielli2004,Welberry1980}, generated from the integer lattice by independently displacing each particle via a uniform distribution in $[-\delta, \delta]$, and
\item
Poisson point distribution configurations.
\end{enumerate}
\noindent We note that the perturbed lattices become identical to the Poisson point distribution configurations if $\delta=N/2$ under the periodic boundary condition.

We denote $M$ constraints, used in the inversion task, by $E_i = 0$ for $i=1,2,\cdots,M$.
Starting from the origin in the Fourier space, we skip the first $N_k$ wavenumbers and constrain the collective coordinates at the next $\floor{M/2}$ wavenumbers:
\begin{equation}\label{eq:constraints1}
E_ i \equiv
\begin{cond}
\Real{\colT{k_{N_k + m}} -\col{k_{N_k +m}}} ,& i=2m-1~~ (i<M) \\
\Imag{\colT{k_{N_k + m}} -\col{k_{N_k +m}}} ,& i=2m,~~~~~~ (i\leq M)
\end{cond}
\end{equation} 
where $\floor{x}$ is the floor function,  $m\in \N$, and $\Real{x}$ and $\Imag{x}$ represent the real and the imaginary parts of a complex number $x$, respectively.
Thus, if $M$ is an even number, both the real and the imaginary parts of collective coordinates at $M/2$ consecutive wavenumbers are constrained.
If $M$ is an odd number, we prescribe the last term $E_M$ via two conditions, each of which is concerning either the real or the imaginary parts of a target collective coordinate as follows:
\begin{numcases}{E_M=}
\Real{\colT{k_{N_k + \ceil{M/2}}} -\col{k_{N_k +\ceil{M/2}}}},  \label{cond:real} \\
\Imag{\colT{k_{N_k + \ceil{M/2}}} -\col{k_{N_k +\ceil{M/2}}}}, \label{cond:imag}
\end{numcases}
where $\ceil{x}$ is the ceiling function.
Table. \ref{table:ex_constraints} lists some examples of constraints.
\begin{table}[ht]
\caption{Examples of constraints $E_i$ for corresponding shorthand notations. We note that when $M$ is an even number, the real condition \eqref{cond:real} and the imaginary condition \eqref{cond:imag} give the identical collective-coordinate constraints.\label{table:ex_constraints}}
\begin{footnotesize}
\begin{tabular}{c|c|c|c|c}
\hline 
• & $E_1$ & $E_2$ & $E_3$ & $E_4$ \\ 
\hline 
$N_k = 0$ and  $M=4$ & $\Real{\colT{k_1} -\col{k_1}}$ & $\Imag{\colT{k_1} -\col{k_1}}$ & $\Real{\colT{k_2} -\col{k_2}}$ & $\Imag{\colT{k_2} -\col{k_2}}$ \\ 
\hline 
$N_k = 1$ and $M=4$ & $\Real{\colT{k_2} -\col{k_2}}$ & $\Imag{\colT{k_2} -\col{k_2}}$ & $\Real{\colT{k_3} -\col{k_3}}$ & $\Imag{\colT{k_3} -\col{k_3}}$ \\ 
\hline
$N_k = 0$,  $M=3$, and & \multirow{2}{*}{$\Real{\colT{k_1} -\col{k_1}}$} & \multirow{2}{*}{$\Imag{\colT{k_1} -\col{k_1}}$} & \multirow{2}{*}{$\Real{\colT{k_2} -\col{k_2}}$} & \multirow{2}{*}{$\cdot$} \\
the real condition \eqref{cond:real} &  &  &  & \\ 
\hline 
$N_k = 0$,  $M=3$, and & \multirow{2}{*}{$\Real{\colT{k_1} -\col{k_1}}$} & \multirow{2}{*}{$\Imag{\colT{k_1} -\col{k_1}}$} & \multirow{2}{*}{$\Imag{\colT{k_2} -\col{k_2}}$} & \multirow{2}{*}{$\cdot$} \\
the imaginary condition \eqref{cond:imag} &  &  &  & \\ 
\hline 
\end{tabular} 
\end{footnotesize}
\end{table}

\section{Numerical Method}\label{sec:numerical methods}

Given a target configuration $\vect{R}^N$ of $N\geq3$, we take $M$ constraints from its collective coordinates, and numerically find solution configurations $\vect{r}^N$ via a modified ``collective-coordinate optimization technique'' \cite{ Zhang2015, Zhang2015b,Uche2004, Uche2006, Zhang2017c} that was initially designed to generate disordered classical point configurations, such as stealthy ground states \cite{Torquato2015_stealthy, Zhang2015,Batten2008}, and the perfect-glass model \cite{Zhang2016b}.
The detailed procedure is described as follows:

\begin{enumerate}
\item \label{step1}
Starting from a random initial configuration $\{r^{(0)}_i \}_{i=1}^N$ of $N$ particles, numerically search for an energy-minimizing configuration $\vect{r}^N\equiv\{\vect{r}_i\}_{i=1}^N$ for the following potential energy,
\begin{eqnarray}\label{eq:Potential energy}
\fl\fn{\Phi}{\vect{r}^N; \vect{R}^N} &\equiv \sum_{l=1}^M \abs{\fn{E_l}{\vect{r}^N; \vect{R}^N}}^2 \nonumber \\
\fl &= 
\begin{cond}
\sum_{l=N_k+1}^{M/2+N_k} \abs{\colT{k_l} - \col{k_l}}^2  , & M \mathrm{~is~even} \\
\sum_{l=N_k+1}^{\floor{M/2}+N_k} \abs{\colT{k_l} - \col{k_l}}^2 + \abs{\fn{E_M}{\vect{r}^N;\vect{R}^N} }^2 , & M \mathrm{~is~odd}.
\end{cond}
\end{eqnarray}
The $j$th component of its gradient is given by
\begin{eqnarray}
\fl \fn{F_j}{\vect{r}^N;\vect{R}^N} &\equiv -\fn{\frac{\partial \Phi}{\partial r_j}}{\vect{r}^N; \vect{R}^N}  \nonumber \\
\fl & = \begin{cond}
\sum_{l=N_k +1} ^{M/2 +N_k} 2k_l \Imag{\left(	\col{k_l}-\colT{k_l} \right)e^{ik_l~ r_j} }  , & M \mathrm{~is~even} \\
\sum_{l=N_k +1} ^{\floor{M/2} +N_k} 2k_l \Imag{\left(\col{k_l}-	\colT{k_l}\right)e^{ik_l ~ r_j} } - 2E_M \pder{E_M}{r_j}, & M \mathrm{~is~odd},
\end{cond} 
\end{eqnarray}
where $E_l$ is defined by \eqref{eq:constraints1}, and for an odd number $M$, $E_M$  is defined by one of two conditions \eqref{cond:real} and \eqref{cond:imag}.
This configuration is called a ``solution'' if $\fn{\Phi}{\vect{r}^N; \vect{R}^N}<\epsilon_E$ for a specified small tolerance $\epsilon_E$.

\item \label{step2}
Test if this solution $\vect{r}^N$ agrees with the target configuration $\vect{R}^N$ or other solutions found previously within another small tolerance $\epsilon_X$, i.e., $\max_{i=1}^N\{\min_{j=1}^N \{\abs{r_i-R_j} \} \}<\epsilon_X$.
If they agree, then $\vect{r}^N$ is deemed to be identical to one of the previous solutions, and we increase the solution's count.
Otherwise, we record $\vect{r}^N$ as a new solution.

\item \label{step3}
Repeat the steps \ref{step1}-\ref{step2} for $N_I$ random initial configurations.
\item \label{step7}
Repeat the steps \ref{step1}-\ref{step3} for $N_T$ different target configurations.
\end{enumerate}
Roughly speaking, the potential \eqref{eq:Potential energy} represents a ``deviation'' or numerical error of a solution configuration from the target configuration in terms of given collective-coordinate constraints.
In step \ref{step1}, we mainly use two different optimization algorithms: the low-storage BFGS (L-BFGS) algorithm \cite{L-BFGS_1980,NLopt_package} with the MINOP algorithm \cite{Dennis1979, Zhang2015}, and the steepest descent algorithm \cite{GNU_manual}. %, the Polak-Ribiere conjugate gradient algorithm \cite{GNU_optimizor, Grippo1997}, and local gradient descent algorithm \cite{Zhang2015}.
%To avoid dependence on an initial configuration $\{r_i^{(0)} \}_{i=1}^N$ and a target configuration $\vect{R}^N$, 
We repeat this inversion task for many distinct initial configurations $\{r_i^{(0)} \}_{i=1}^N$s and target configurations $\vect{R}^N$s.
Unless stated otherwise, we use parameters as follows: $N_I = 1~000$, $N_T = 1~000$, and $\epsilon_X = 10^{-6}$.

For all numerically distinct solutions $\{\vect{r}^N \}$ of a target configuration $\vect{R}^N$, the \textit{trivial solution} refers to the one that is identical to the target ($\vect{r}^N = \vect{R}^N$), while \textit{nontrivial} solutions refer to the others ($\vect{r}^N \neq \vect{R}^N$).

\section{Results for $N\leq 3$}\label{sec:small systems}
Here, we theoretically and numerically investigate solutions for small target configurations.

\subsection{$N=1$}\label{sec:N=1}
\begin{figure}[ht]
\begin{center}
\includegraphics[width=0.4\textwidth]{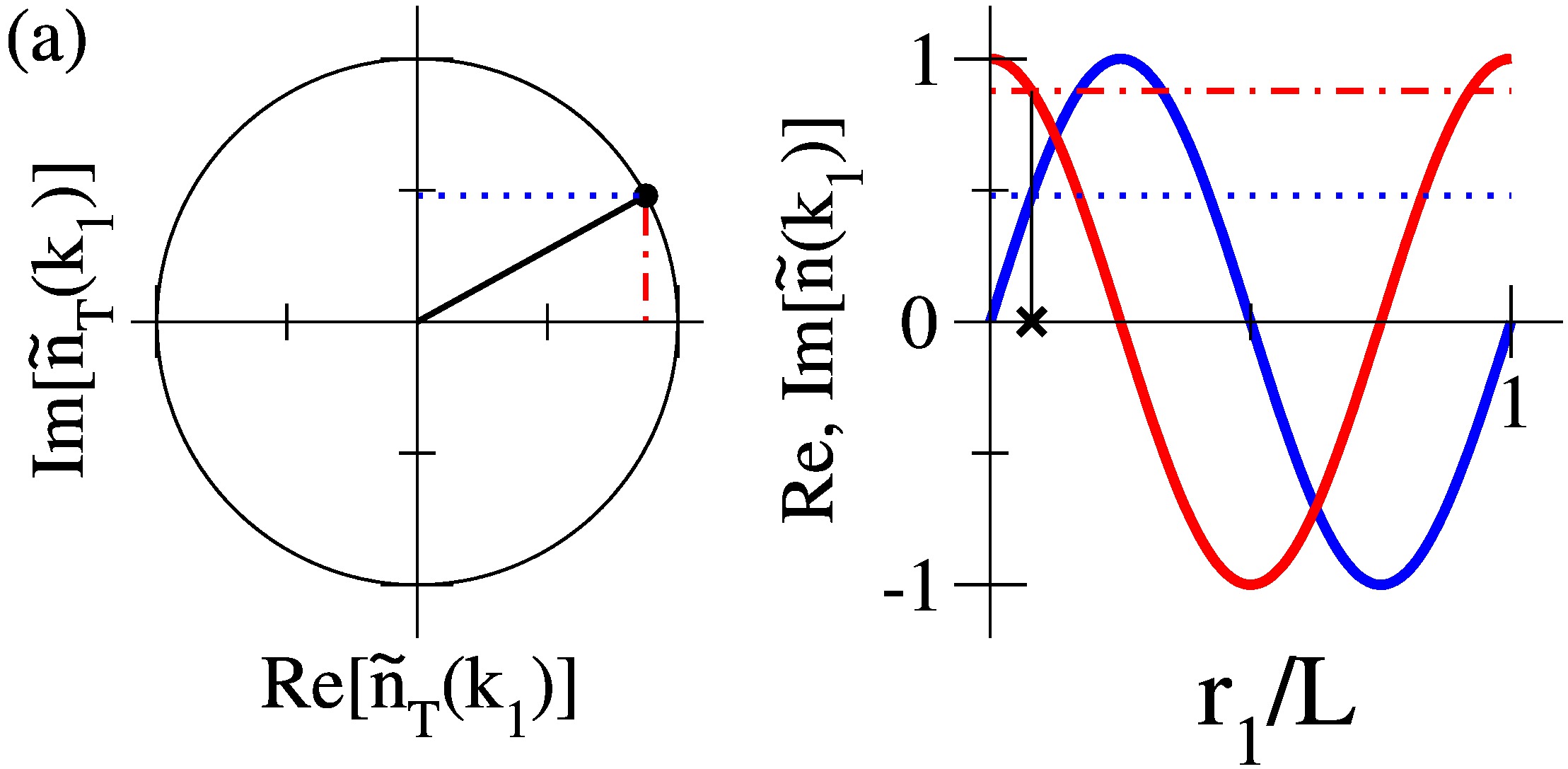}
\hspace{10pt}
\includegraphics[width=0.4\textwidth]{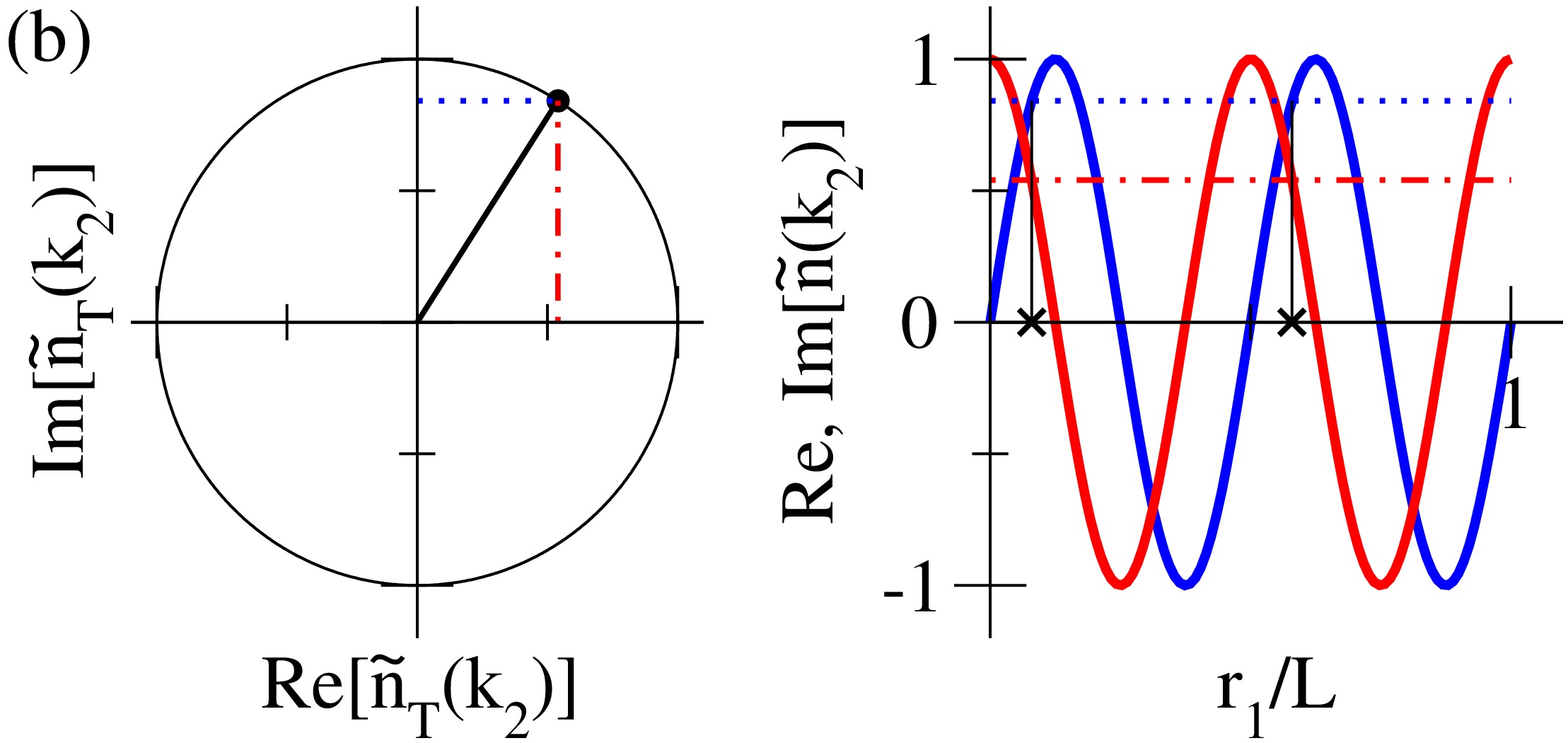}
\end{center}
\caption{Illustrations for solutions of the inversion problem for a single-particle target configuration. 
(a) Cases with $N_k=0$ and $M=2$. When $\colT{k_1}$ is given as constraints (left), both its real and imaginary parts are required for a unique solution; see the cross ($\times$) mark in the right panel.
Red and blue lines represent the real and the imaginary parts of $\col{k_1}$ of a solution, respectively.
(b) Cases with $N_k=1$ and $M=2$. When $\colT{k_2}$ is given, we have two solutions.  \label{fig:N_1 example}}
\end{figure}

For a single-particle configuration, $\col{k_1} = e^{-i2\pi r_1/L}$ is a one-to-one function from $r_1 \in [0,L)$ onto the unit circle on the complex plane, i.e., $\left\{ z\in \C : \abs{z}=1 \right\}$.
Thus, it is straightforward to show that there is a unique solution, given constraints $\colT{k_1}$ that correspond to the cases of $N_k=0$, and $M=2$.
Equivalently, collective coordinates at larger wavenumbers can be expressed in terms of $\colT{k_1}$, i.e., $\colT{k_m} = {\colT{k_1}}^m$.
On the other hand, cases of $N_k=0$ and $M=1$, i.e., a single constraint of either $\Real{\colT{k_1}}$  or $\Imag{\colT{k_1}}$, give two solutions; see figure \ref{fig:N_1 example}(a).
Thus, we need at least two constraints ($M=2$) for the unique inversion of a single-particle configuration.

We note that $\colT{k_1}$ is the minimal set of constraints for single-particle systems.
This is because when $m>1$, $\colT{k_m}$ is no longer a one-to-one function from $r_1 \in [0,L)$ onto the unit circle on $\C$, and thus cases with $N_k =m$ and $M=2$ for $m>1$ give $m$ distinct solutions; see figure \ref{fig:N_1 example}(b).

\subsection{$N=2$}\label{sec:N=2}

\begin{figure}[h]
\begin{center}
\subfloat[$\vect{R}^2 /L = (0.1, 0.1)$]{\includegraphics[width=0.3\textwidth]{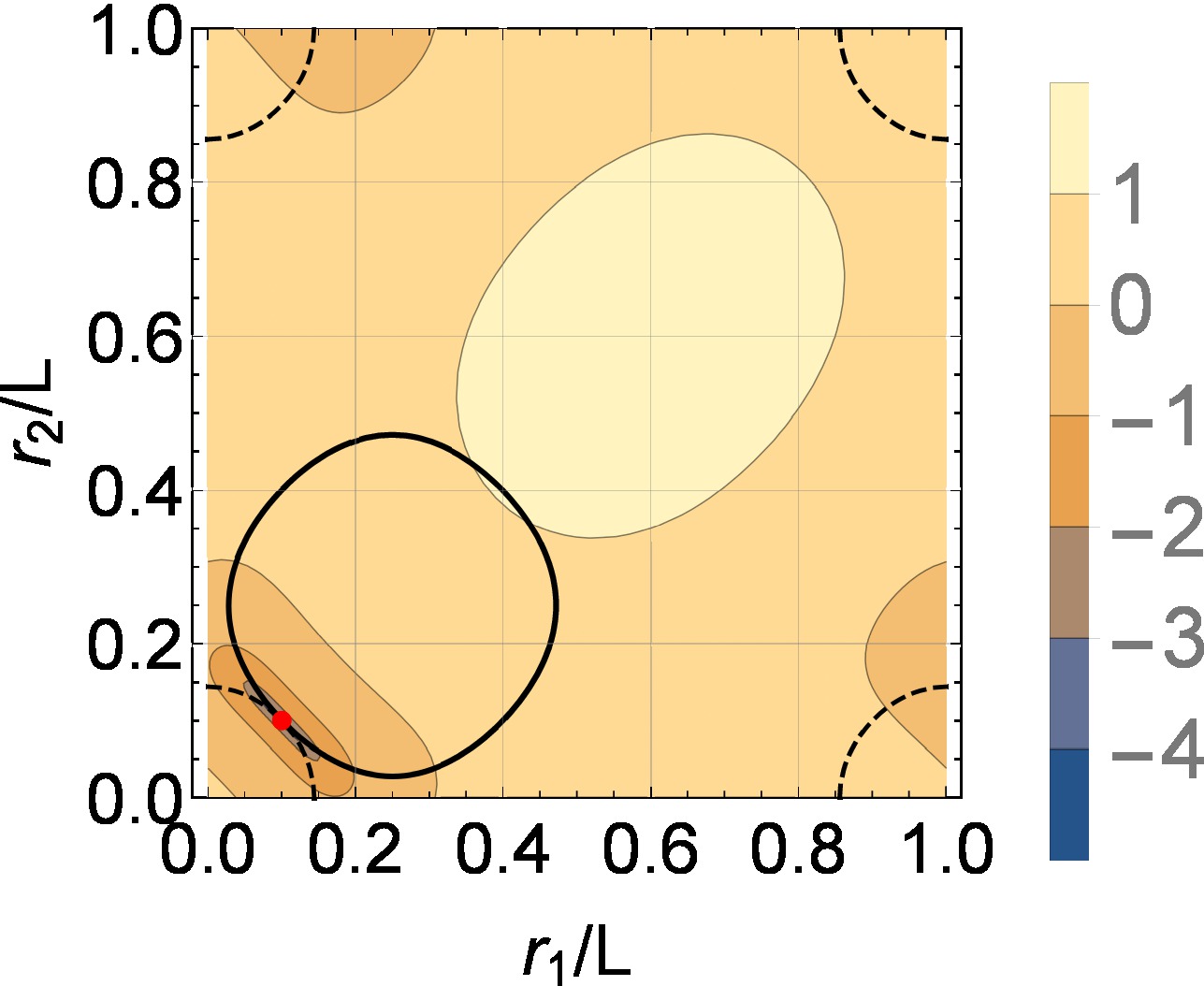}}
\hspace{5pt}
\subfloat[$\vect{R}^2 /L = (0.1, 0.3)$]{\includegraphics[width=0.3\textwidth]{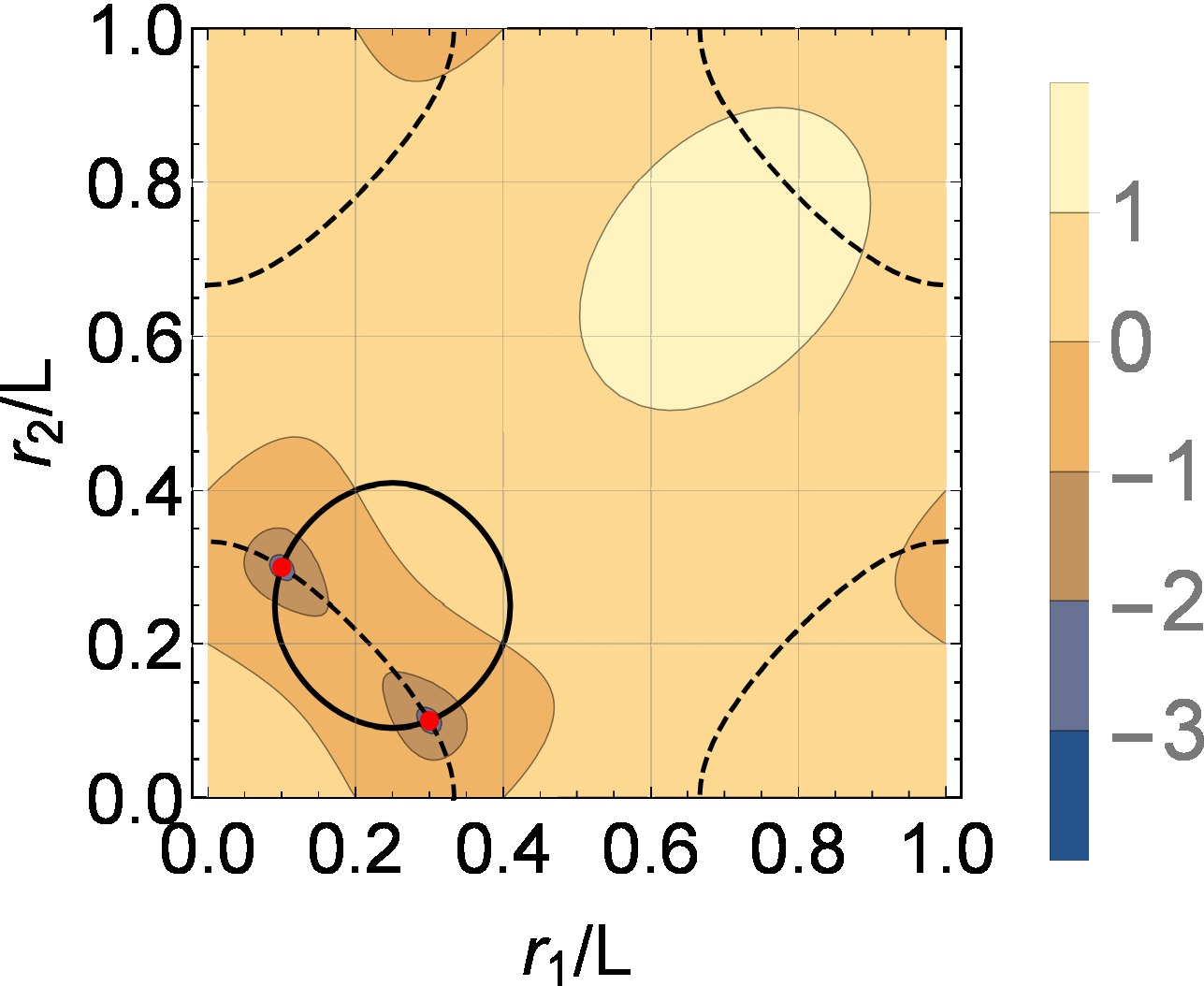}}
\hspace{5pt}
\subfloat[$\vect{R}^2 /L = (0.1, 0.6)$]{\includegraphics[width=0.3\textwidth]{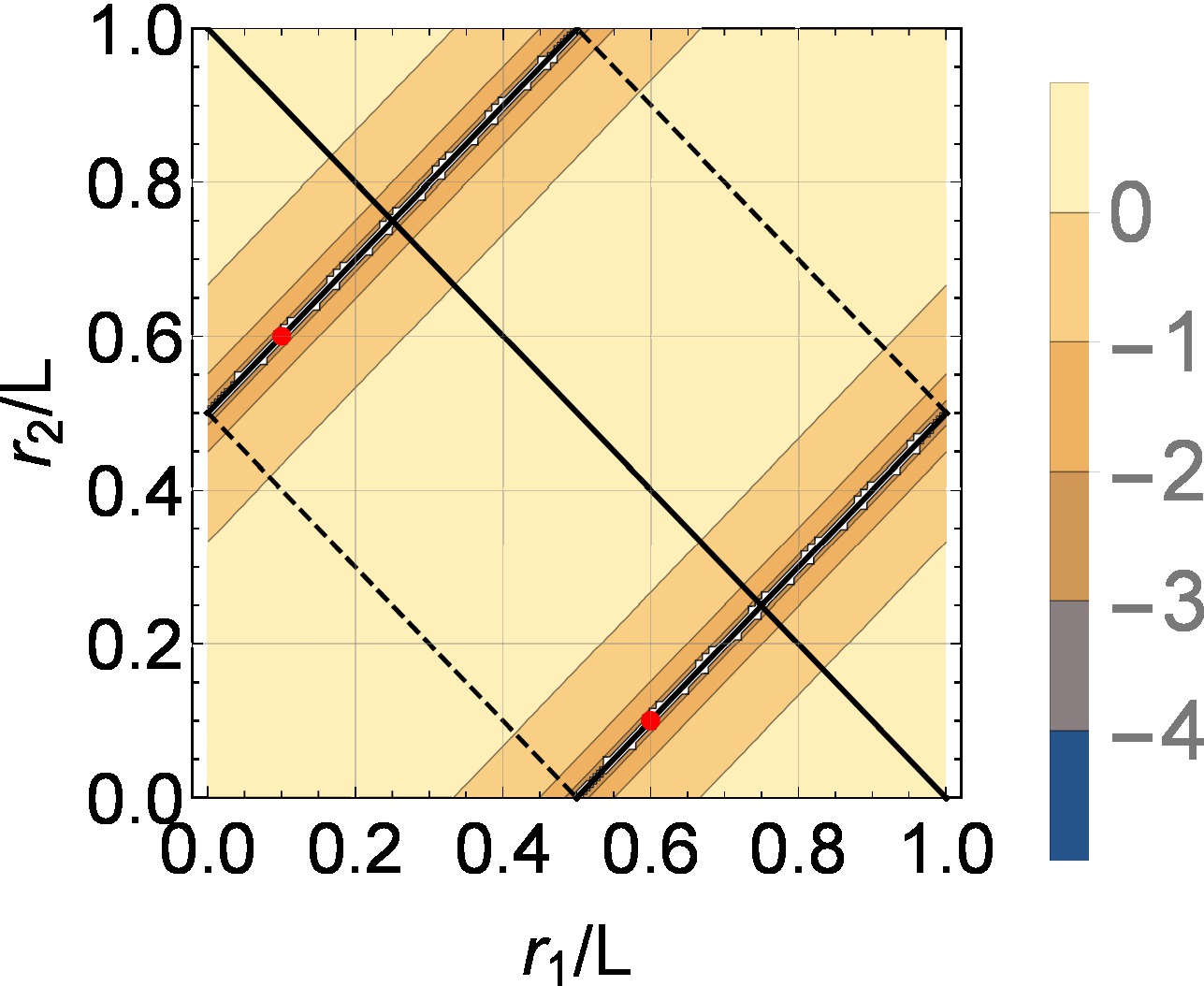}}

\subfloat[$\vect{R}^2 /L = (0.1, 0.55)$]{\includegraphics[width=0.3\textwidth]{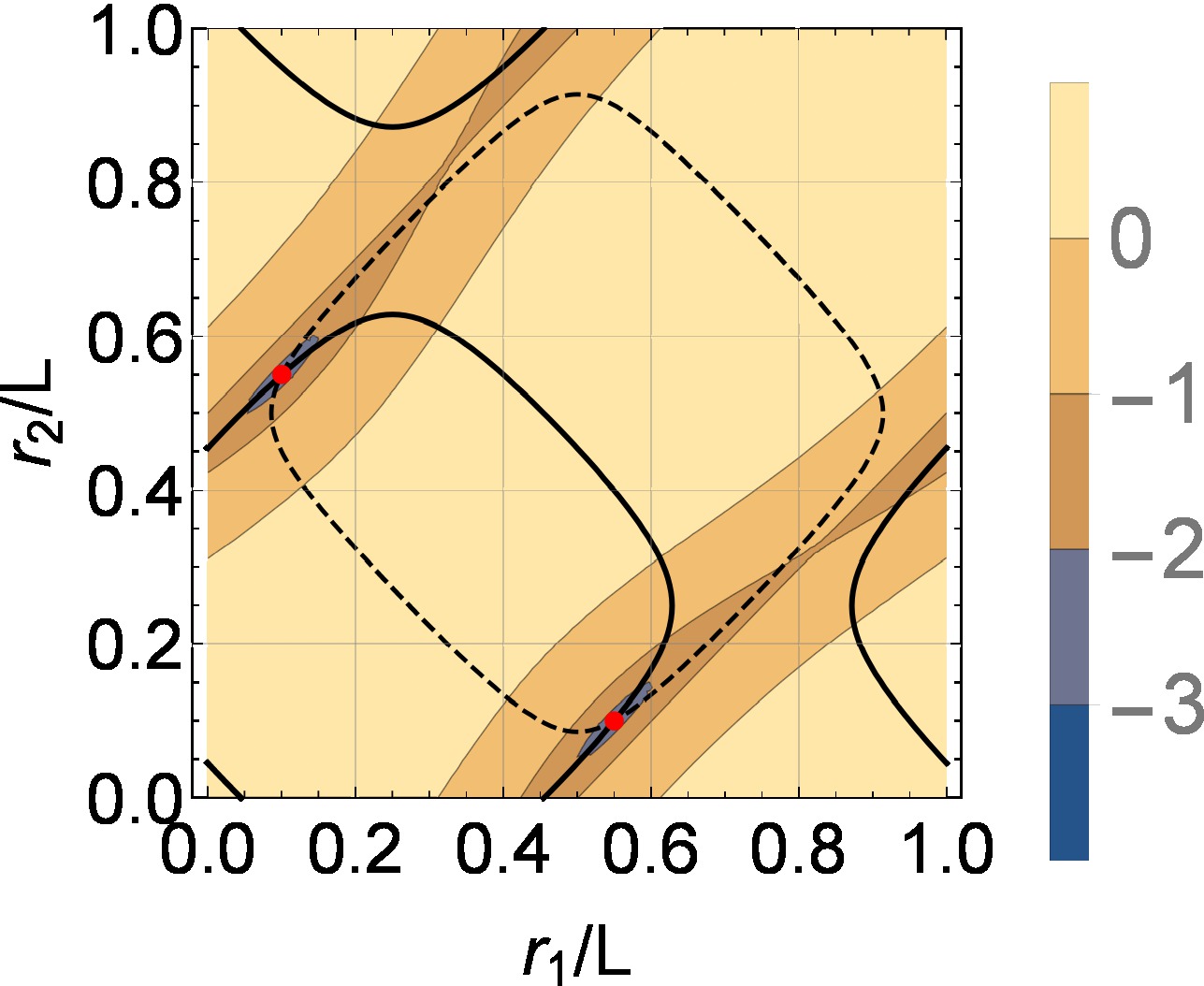}}
\hspace{5pt}
\subfloat[$\vect{R}^2 /L = (0.1, 0.59)$]{\includegraphics[width=0.3\textwidth]{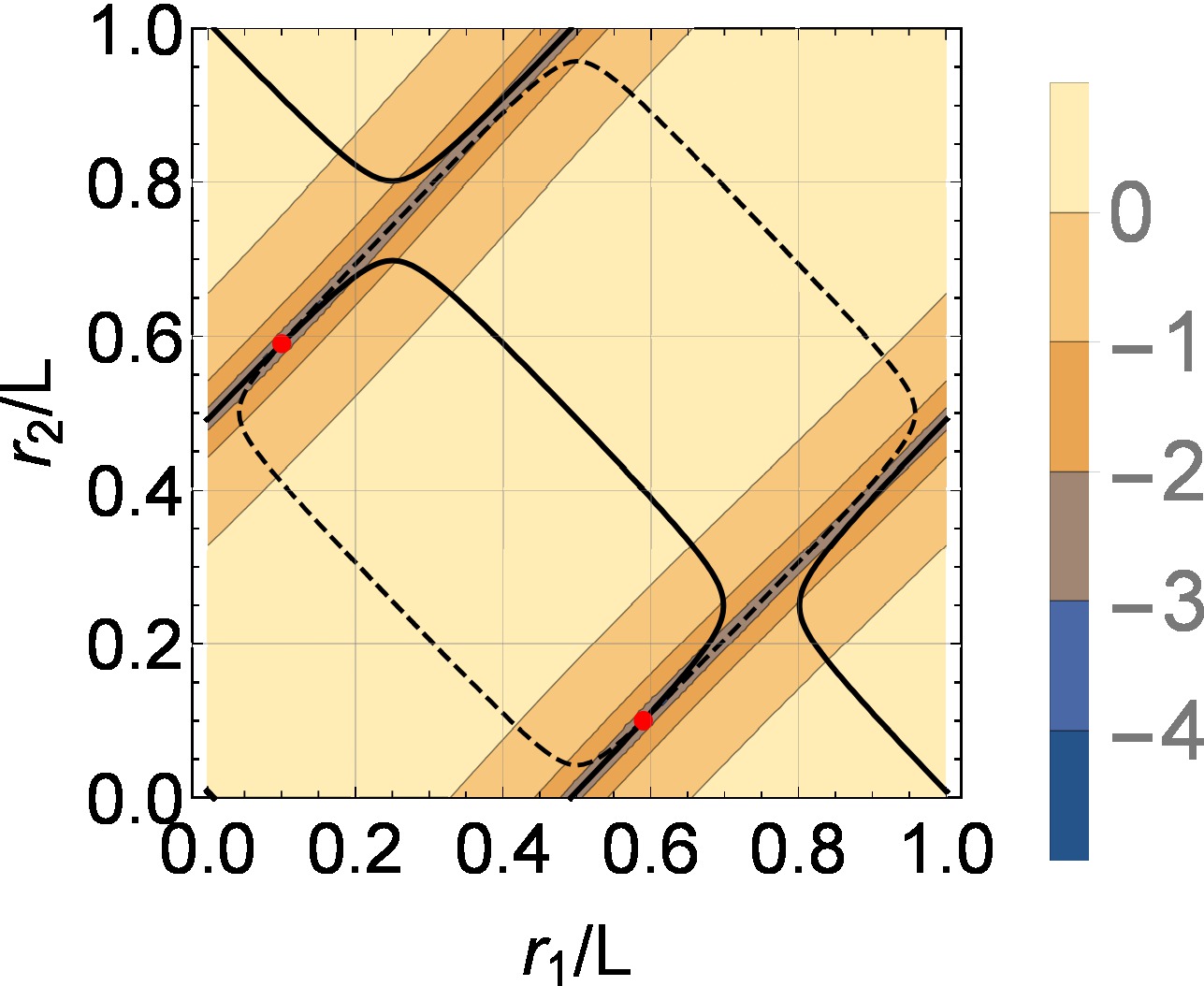}}
\hspace{5pt}
\subfloat[$\vect{R}^2 /L = (0.1, 0.65)$]{\includegraphics[width=0.3\textwidth]{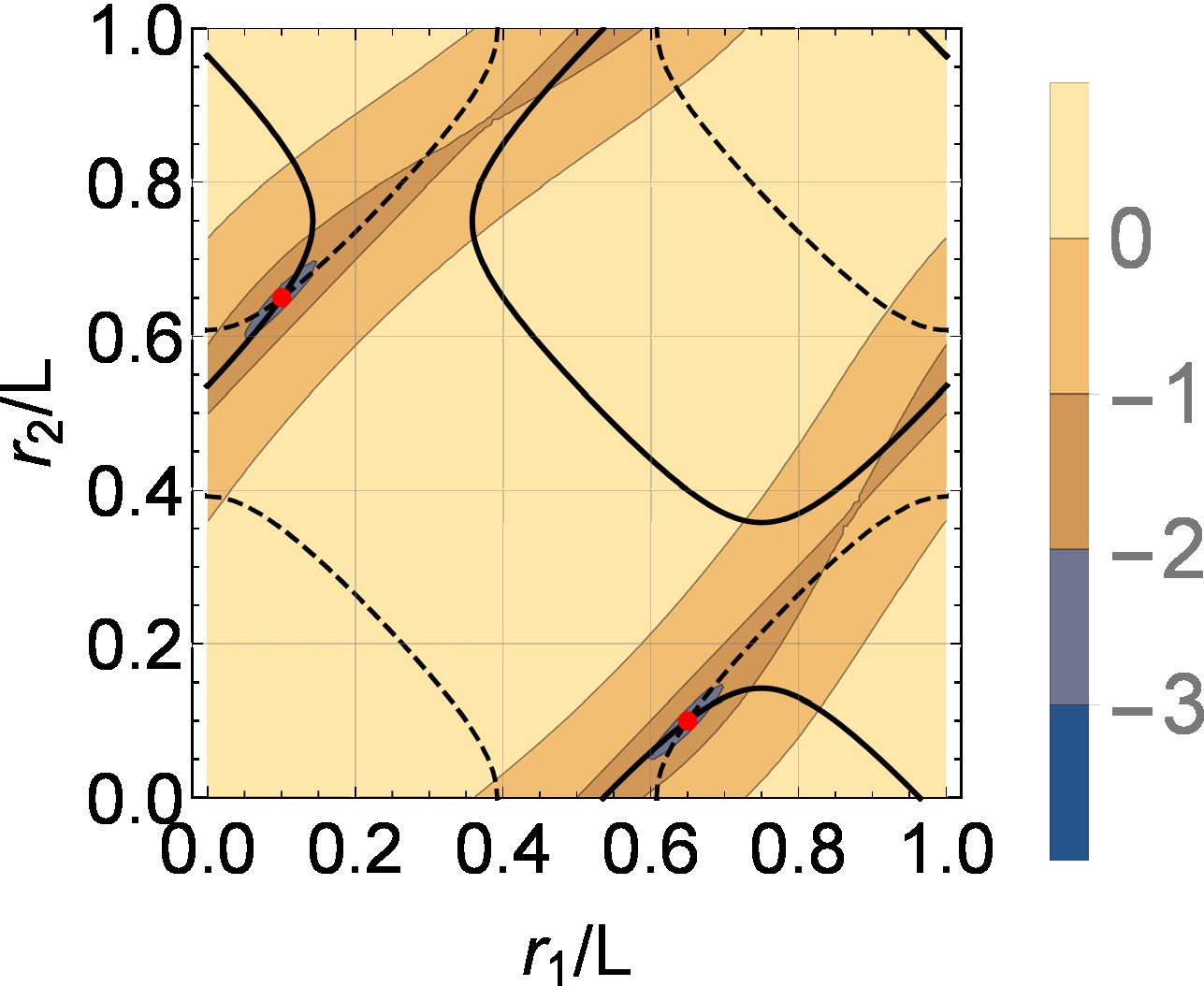}}
\end{center}
\caption{Graphical solutions of \eqref{eq:2pt_constraints} for given respective target configurations.
In each panel, black solid lines and dashed ones represent solutions of the ``real'' and the ``imaginary'' parts of  \eqref{eq:2pt_constraints}, respectively.
Contour plots depict potential energy landscape [i.e., $\log_{10}(\fn{\Phi}{\vect{r}^2;\vect{R}^2})$] for each target configuration.
Solutions (intersections of solid and dashed lines) are unique and identical to the target configuration (red dots), unless it is the integer lattice (i.e., $\abs{R_1-R_2} = L/2$) as shown in (c).
Otherwise, there are infinitely many solutions, and one needs additional constraint $\colT{k_2}$ for unique solutions.
\label{fig:graphical solution_2pt}}
\end{figure}

Using graphical solutions, one can straightforwardly show a single constraint ($N_k=0$ and $M=1$) gives infinitely many solutions; see one of the solid or dashed lines in figure \ref{fig:graphical solution_2pt}.
However, figure \ref{fig:graphical solution_2pt} also immediately shows that the following equation ($N_k=0$ and $M=2$)
\begin{equation}\label{eq:2pt_constraints}
\colT{k_1} = e^{-i2\pi r_1/L}+ e^{-i2\pi r_2/L},
\end{equation}
and it yields a unique solution under exchanges of particle indices, as follows:
\begin{eqnarray}
e^{-i2\pi r_1/L} & = \frac{\colT{k_1}}{2}\left(1\pm i \sqrt{\frac{4}{\abs{\colT{k_1}}^2}-1}\right) \label{eq:n_1D2prts_1}\\
e^{-i2\pi r_2/L} & = \frac{\colT{k_1}}{2}\left(1\mp i \sqrt{\frac{4}{\abs{\colT{k_1}}^2}-1}\right),\label{eq:n_1D2prts_2}
\end{eqnarray}
if $\colT{k_1} \neq 0$, or equivalently, $\abs{R_1-R_2}\neq 0.5 L$.
Otherwise, the periodic image of the target configuration becomes the integer lattice, and all of translated lattices are solutions of \eqref{eq:2pt_constraints}, i.e., there are infinitely many solutions, as shown in figure \ref{fig:graphical solution_2pt}(c).

%the periodic image of the target configuration becomes the integer lattice, and the expressions \eqref{eq:n_1D2prts_1} and \eqref{eq:n_1D2prts_2} become singular.
%Instead, as shown in figure \ref{fig:graphical solution_2pt} (c), all of translated lattices are solutions of \eqref{eq:2pt_constraints}, i.e., there are infinitely many solutions.

If the target configuration is the integer lattice, in order to obtain a unique solution, the collective coordinate at the first Bragg peak [i.e., $\colT{k_2}$] should be additionally specified, which corresponds to the cases with $N_k=0$ and $M=4$.
Then, the unique solution is 
\begin{eqnarray}
e^{-i2\pi r_1/L} &= \frac{1}{2} \left(\colT{k_1} \pm \sqrt{2{\colT{k_2}}^2 -{\colT{k_1}}^2}	\right),\\
e^{-i2\pi r_2/L} &= \frac{1}{2} \left(\colT{k_1} \mp \sqrt{2{\colT{k_2}}^2 -{\colT{k_1}}^2}	\right).
\end{eqnarray}
This is because the collective coordinate at the first Bragg peak provides the center of mass of this lattice configuration.

We note that the constraint $\colT{k_2}$ alone (i.e., $N_k=1$ and $M=2$) cannot be uniquely inverted into particle coordinates.
It can be  straightforwardly shown that there exist at least four distinct solutions, i.e., $(r_1,r_2) = \vect{a} + (R_1,R_2)$,
where $\vect{a}/L = (0,0),~(0,1/2),~(1/2,0),$ and $(1/2,1/2)$.
By the same analysis, one can identify there are at least $m^2$ distinct solutions if only $\colT{2\pi m/L}$ is given.
Therefore, we can conclude that for a two-particle configuration that is not the integer lattice, the minimal set of constraints for a unique solution is $\{\colT{k_1} \}$.

{\bf Remarks}
\begin{enumerate}
\item For a configuration of particle number $N>1$, Fan, \textit{et al.} \cite{Fan1991} proved that $\col{k_m} = 0$ for $m=1,\cdots,\floor{\frac{N}{2}}$ is a sufficient and necessary condition for the configuration to be the integer lattice or its translations.
Thus, if one inverts collective coordinates at the $\ceil{N/2}$ smallest wavenumbers of the integer lattice, its solutions are inevitably degenerated with a translational degree of freedom; see figure \ref{fig:graphical solution_2pt} (c) for example.

\end{enumerate}

\subsection{$N=3$}\label{sec:N=3}
In the previous sections, we show that there is a unique solution in the inversion procedure with parameters $N_k = 0$ and $M = \ceil{N/2}$, unless the target configuration is a pathological case (i.e., either the integer lattice or its translations).
Otherwise, there are infinitely many solutions.
It implies that there would be a sudden transition in the number of distinct solutions varying with the type of target configurations.
For this reason and simplicity in analysis, our target configurations are restricted here to perturbed lattices that can continuously interpolate between the integer lattice to Poisson configurations via the displacement parameter $\delta$; see section \ref{sec:def}.

\begin{figure}[ht]
\begin{center}
\subfloat[]{\includegraphics[width=0.36\textwidth]{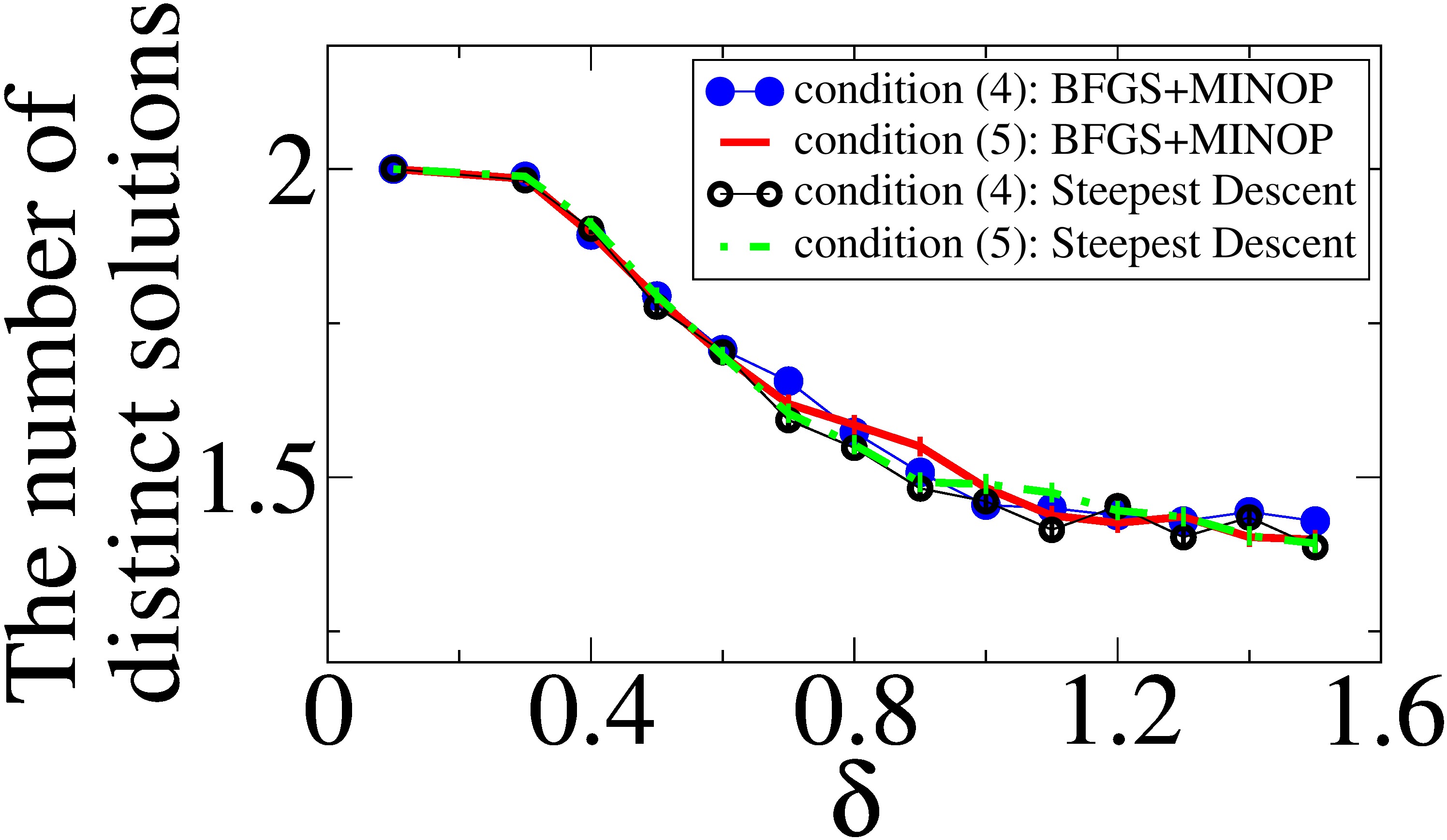}}
\hspace{10pt}
\subfloat[]{\includegraphics[width = 0.48\textwidth]{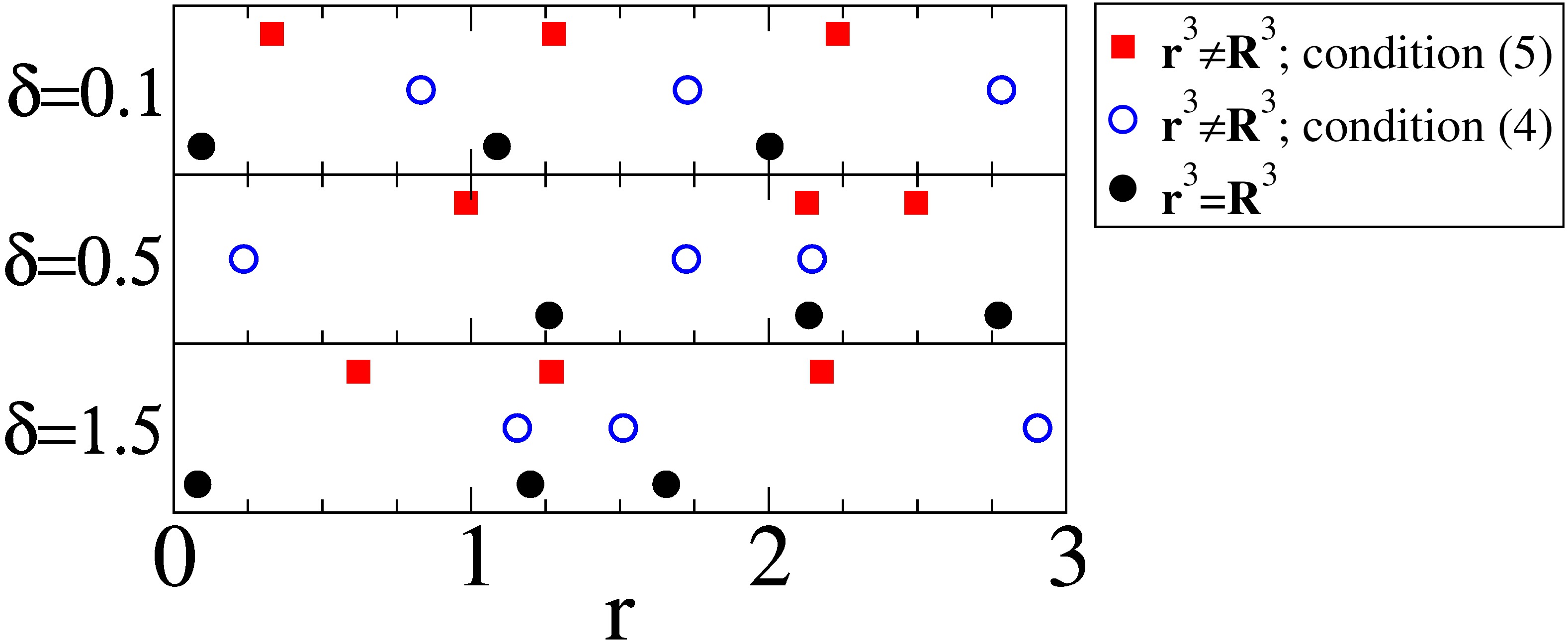}}
\end{center}
\caption{Numerical results of the inversion procedure for three-particle perturbed lattices in cases with $N_k=0$ and $M=3$.
(a) The average number of distinct solutions per a target configuration.
Two different optimization algorithms (BFGS+MINOP and the steepest descent) and two constraint conditions [the real \eqref{cond:real} and the imaginary \eqref{cond:imag} ones] are used for comparison with the energy tolerance $\epsilon_E =10^{-29}$. 
For any target configuration, the number of distinct solutions is at most two, but the average can vary with the target configurations.
(b) Examples of nontrivial solutions for a given target perturbed lattice with various displacements $\delta$.
Nontrivial solutions by the real \eqref{cond:real} or the imaginary \eqref{cond:imag} conditions, respectively, are different from each other, and are not translations of the target.}
\label{fig:3pt solutions} 
\end{figure}

For a perturbed lattice, its particle coordinates are described as $r_i = (i-1) + N \delta_i$ for $i=1,\cdots, N$.
Assuming weak perturbations (i.e., $\abs{\delta_i} \ll 1$) for $N=3$, collective-coordinate constraints can be approximated up to the second order of displacements;
\begin{equation}
\Real{\col{k_m}} \approx 
\begin{cond}
3-2(m\pi)^2 ({\delta_1}^2 +{\delta_2}^2 + {\delta_3}^2), & m = 3i \\
\sqrt{3} m \pi  (\delta_2-\delta_3) +(m\pi)^2 (-2{\delta_1}^2 +{\delta_2}^2 +{\delta_3}^2), & m =3i+1 \\
-\sqrt{3} m \pi  (\delta_2-\delta_3) +(m\pi)^2 (-2{\delta_1}^2 +{\delta_2}^2 +{\delta_3}^2), & m =3i+2
\end{cond}  \label{eq:Cn_N=3}
\end{equation}
\begin{equation}
\Imag{\col{k_m}} \approx 
\begin{cond}
2 m\pi \left(\delta_1 +\delta_2 +\delta_3	\right), & m = 3i \\
m\pi \left(2\delta_1 -\delta_2-\delta_3	\right) + \sqrt{3}(m\pi)^2 \left(	{\delta_2}^2 -{\delta_3}^2\right), &m = 3i+1 \\
m\pi \left(2\delta_1 -\delta_2-\delta_3	\right) - \sqrt{3}(m\pi)^2 \left(	{\delta_2}^2 -{\delta_3}^2\right), &m = 3i+2 
\end{cond}, \label{eq:Sn_N=3}
\end{equation}
where $i$ represents non-negative integers.

For parameters $N_k=0$ and $M=3$ with the real condition \eqref{cond:real} [or the imaginary one \eqref{cond:imag}], the quadratic approximations \eqref{eq:Cn_N=3} and \eqref{eq:Sn_N=3} yield at most two distinct solutions \eqref{eqs:sol_3prt_1}: the trivial solution ($\vect{r}^3 = \vect{R}^3$), and a nontrivial one ($\vect{r}^3 \neq \vect{R}^3$).
This prediction is consistently observed in numerical results; see figure \ref{fig:3pt solutions}(a).
Thus, the set of numerically  distinct solutions abruptly changes from an uncountably many set into a finite one, as $\delta$ becomes nonzero.
Figure \ref{fig:3pt solutions}(a) also shows that if $\delta$ increases, while the maximal number of numerically distinct solutions remains two, its occurrence decreases regardless of constraint conditions \eqref{cond:real} and \eqref{cond:imag}.

\begin{figure}[ht]
\begin{center}
\includegraphics[width = 0.99\textwidth]{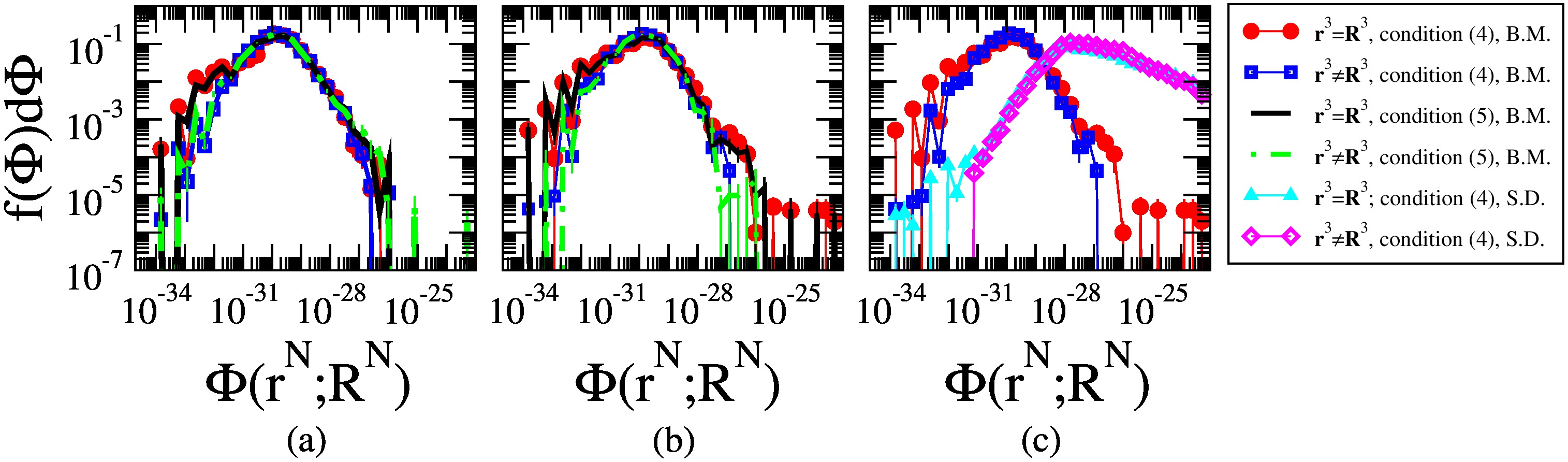}
\end{center}
\caption{Log-log plots of histograms for energy distributions of numerically distinct solutions $\{\vect{r}^3\}$ of a three-particle target configuration $\vect{R}^3$ for parameters $N_k=0$, $M=3$, and $\epsilon_E = 10^{-20}$.
Given a target configuration, there are at most two distinct solutions; a trivial solution and a nontrivial one.
(a-b) Results from two constraint types [i.e., the real condition \eqref{cond:real} and the imaginary condition \eqref{cond:imag}] are compared for two different types of target configurations: (a) perturbed lattices with $\delta=0.1$ and (b) Poisson configurations.
Here, BFGS+MINOP (B.M.) algorithms are used.
(c) For Poissonian target configurations, we compare results from two different optimization algorithms: B.M., and steepest descent (S.D.).
Here, the real condition \eqref{cond:real} is considered.
\label{fig:energy distribution 3prt}}
\end{figure}

In numerical studies, it is important to know how results depend on the optimization algorithms and values of parameters, such as $\epsilon_E$ and $\epsilon_X$.
For this purpose, we investigate the energy distributions of numerical solutions obtained in the parameters of  $N_k=0$ and $M=3$, and various conditions, as shown in figure \ref{fig:energy distribution 3prt}.
From figure \ref{fig:energy distribution 3prt} (a) and (b), we see that given a target configuration, both trivial and nontrivial solutions have qualitatively similar energy profiles, regardless of the real \eqref{cond:real} and the imaginary \eqref{cond:imag} conditions.
Figure \ref{fig:energy distribution 3prt}(c) demonstrates that the energy profiles of numerical solutions vary with optimization algorithms, but for a given algorithm both trivial and nontrivial solutions still have qualitatively similar energy profiles.
Thus, a nontrivial solution cannot be eliminated by lowering the energy tolerance $\epsilon_E$ when $N=M=3$. 
In the rest of this paper, we mainly use the BFGS and MINOP algorithms because the solutions obtained via these algorithms tend to have lower numerical errors than those via the steepest descent method.

For parameters $N_k=0$ and $M=4$, a unique solution can be obtained.
This also can be deduced from the observation in the cases with $N_k=0$ and $M=3$ that given a target configuration, nontrivial solutions, respectively obtained by the real \eqref{cond:real} and the imaginary \eqref{cond:imag} conditions, are numerically distinct; see figure \ref{fig:3pt solutions}(b).
Thus, the common solution from two conditions \eqref{cond:real} and \eqref{cond:imag} should be identical to the target.
The unique solution also can be obtained from the quadratic approximations \eqref{eq:Cn_N=3} and \eqref{eq:Sn_N=3} as follows:
\begin{eqnarray}
\delta_1 & = \frac{1}{12\pi}\left[\frac{6\Imag{2 \colT{k_1} -\colT{k_2}}}{\Real{4\colT{k_1}-\colT{k_2}}} +\Imag{4\colT{k_1} +\colT{k_2}} 	\right] \\
\delta_2 & = \frac{1}{12\pi} \left[\frac{6\Imag{2 \colT{k_1} -\colT{k_2}}}{\Real{4\colT{k_1}-\colT{k_2}}} - \frac{\Real{4\colT{k_1}-\colT{k_2}}}{\sqrt{3}}	\right] \label{eqs:sol_3prt_2}\\
\delta_3 & = \frac{1}{12\pi} \left[\frac{6\Imag{2 \colT{k_1} -\colT{k_2}}}{\Real{4\colT{k_1}-\colT{k_2}}} +\frac{\Real{4\colT{k_1}-\colT{k_2}}}{\sqrt{3}}	\right],
\end{eqnarray}
and thus the minimal set for three-particle systems is (both real and imaginary parts of) collective coordinates at the two smallest wavenumbers.

{\bf Remarks}
\begin{enumerate}
\item For parameters $N_k=0$, $M=3$, and the real condition \eqref{cond:real}, the quadratic approximations \eqref{eq:Cn_N=3} and \eqref{eq:Sn_N=3} give two exact solutions \eqref{eqs:sol_3prt_1}.
While one of the solutions is the same as the target configuration up to some numerical errors, another solution cannot precisely predict the nontrivial solution partly because the nontrivial one is not a perturbed lattice with small displacements.

\item For parameters $N_k=1$ and $M=4$, a unique solution is obtained; see  \eqref{eq:sol_Nk=1_M_4(2)},  \eqref{eq:sol_Nk=1_M_4(3)}, and \eqref{eq:sol_Nk=1_M_4(1)}.

\end{enumerate}

\section{Results for $N>3$}\label{sec:N>3}

\begin{figure}[ht]
\begin{center}
\includegraphics[width=0.9\textwidth]{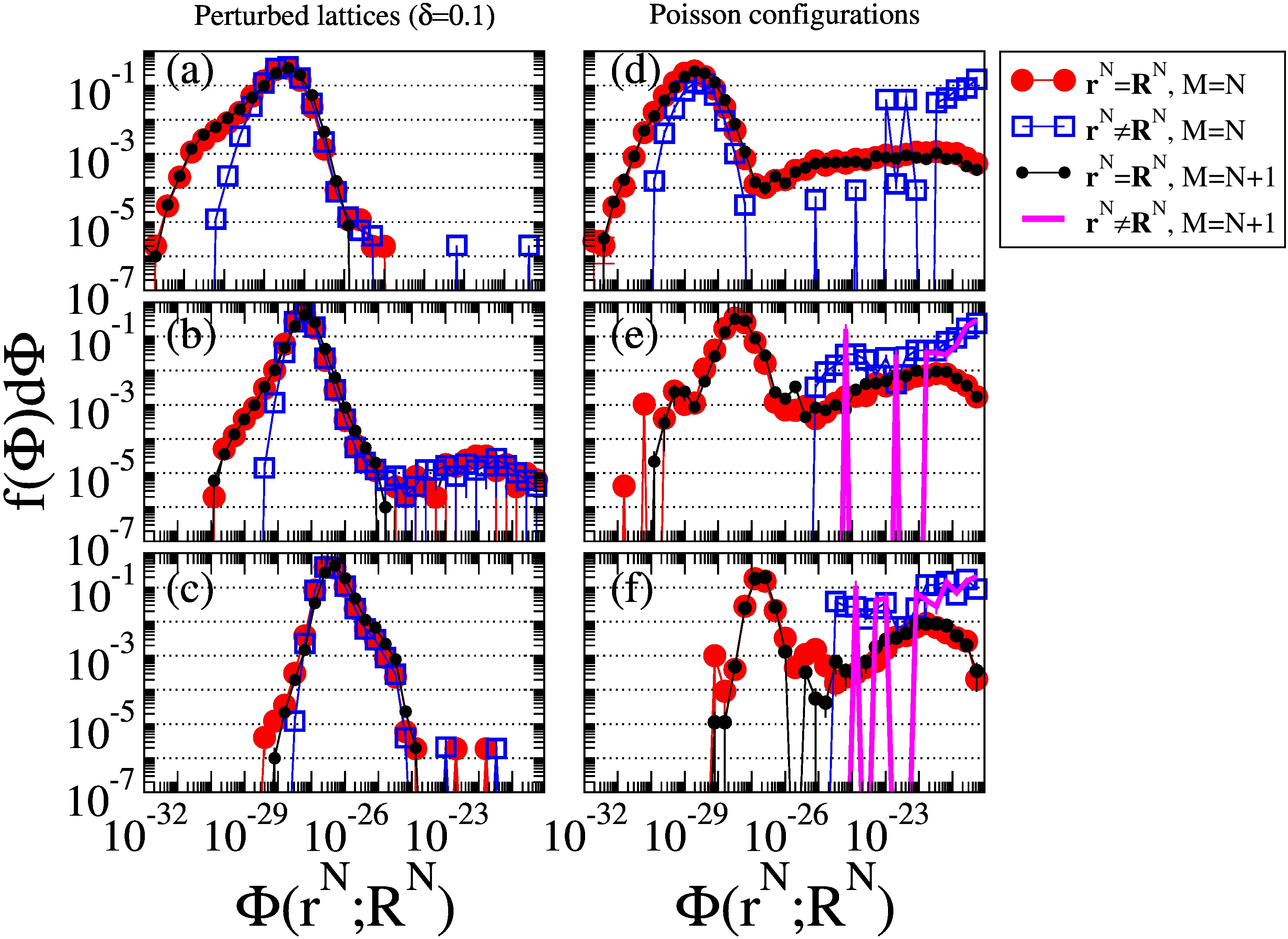}
\end{center}
\caption{Log-log plots of histograms for energy distribution of numerically distinct solutions $\{\vect{r}^N\}$ for odd-number system sizes: $N=9$ (a, d), $19$ (b, e), and $29$ (c, f).
Using the real condition \eqref{cond:real} condition and parameters $N_k=0$ and $\epsilon_E = 10^{-20}$, two types of target configurations are considered: (a-c) perturbed lattices with $\delta=0.1$ and (d-f) Poisson configurations.
When $M=N$, while a target perturbed lattice has a single nontrivial solution ($\vect{r}^N\neq \vect{R}^N$), whose occurrence rate is similar to that of trivial ones, a Poissonian target mainly has the trivial solution but occasionally can have multiple nontrivial solutions.
When $M=N+1$ is an even number, while there is a unique solution for perturbed lattices, there can be more than one solution for a Poisson target configuration in relatively lower occurrence rates. %We note that there is no nontrivial solution when $M=N+1$ in (d).
Even in the latter case, however, the nontrivial solutions can be eliminated by lowering the tolerance $\epsilon_E$ around $10^{-25}$.
\label{fig:energy distributions}}
\end{figure}

\begin{figure}[ht]
\begin{center}
\includegraphics[width = 0.9\textwidth]{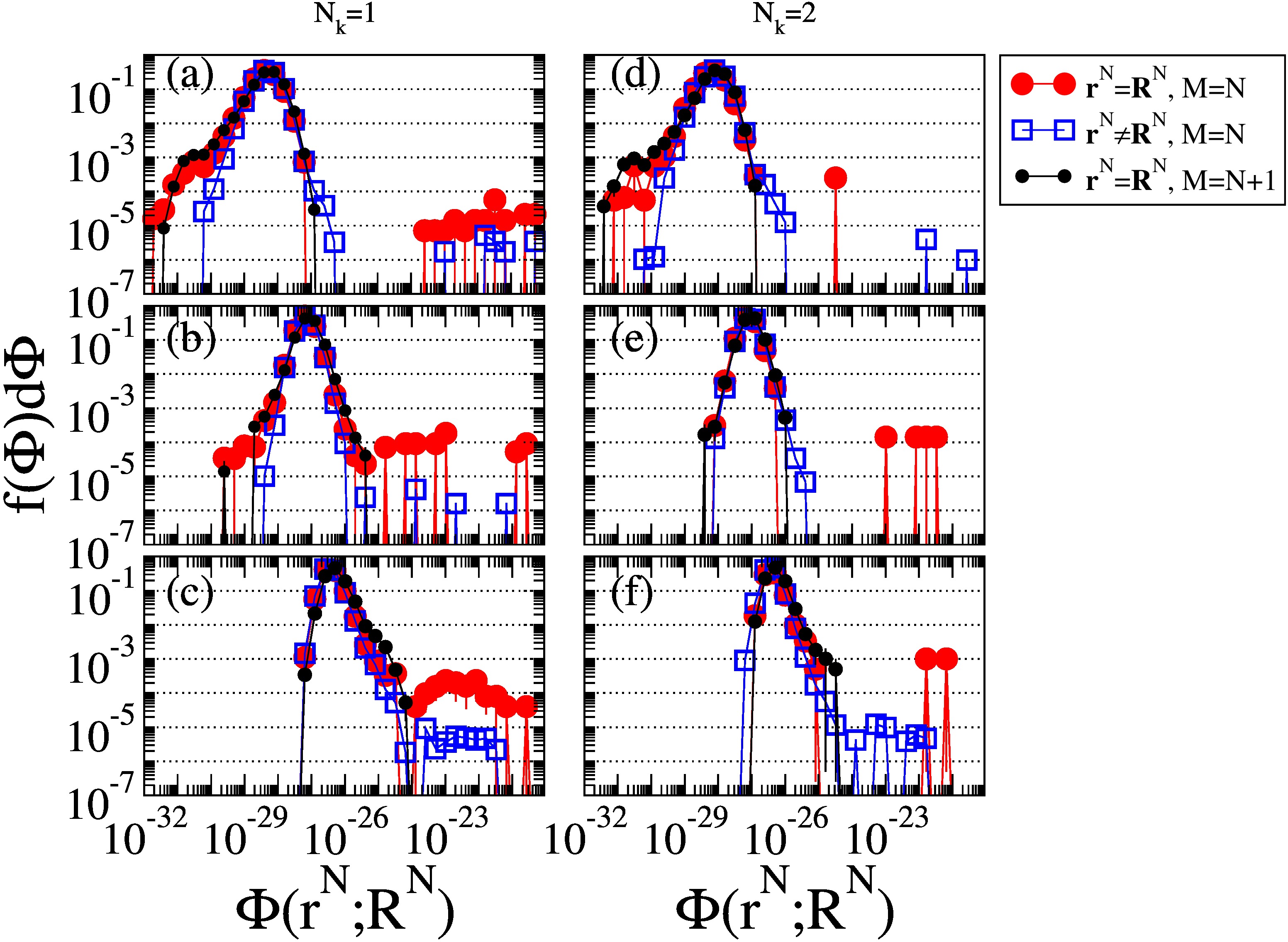}
\end{center}
\caption{Log-log plots of histograms for energy distribution of numerically distinct solutions $\{\vect{r}^N\}$ for $N_k>0$ and odd-number system sizes: $N=9$ (a, d), $19$ (b, e), and $29$ (c, f).
Considering perturbed lattices with $\delta=0.1$ as the target configurations, we search solution configurations under the real condition \eqref{cond:real} and the tolerance $\epsilon_E =10^{-20}$, and via the BFGS+MINOP algorithms.
We note that there is no nontrivial solution with $\fn{\Phi}{\vect{r}^N;\vect{R}^N} < 10^{-20}$ if $N_k > 0$ and $M=N+1$.
\label{fig:energy distribution2}}
\end{figure}

\begin{figure}[ht]
%\subfloat[Perturbed lattices ($\delta=0.1$)]{\includegraphics[height=0.35\textwidth]{sols_PL_01.jpg}}
%\hspace{5pt}
%\subfloat[Poisson configurations]{\includegraphics[height=0.35\textwidth]{sols_Po.jpg}}
\begin{center}
\includegraphics[width = 0.9\textwidth]{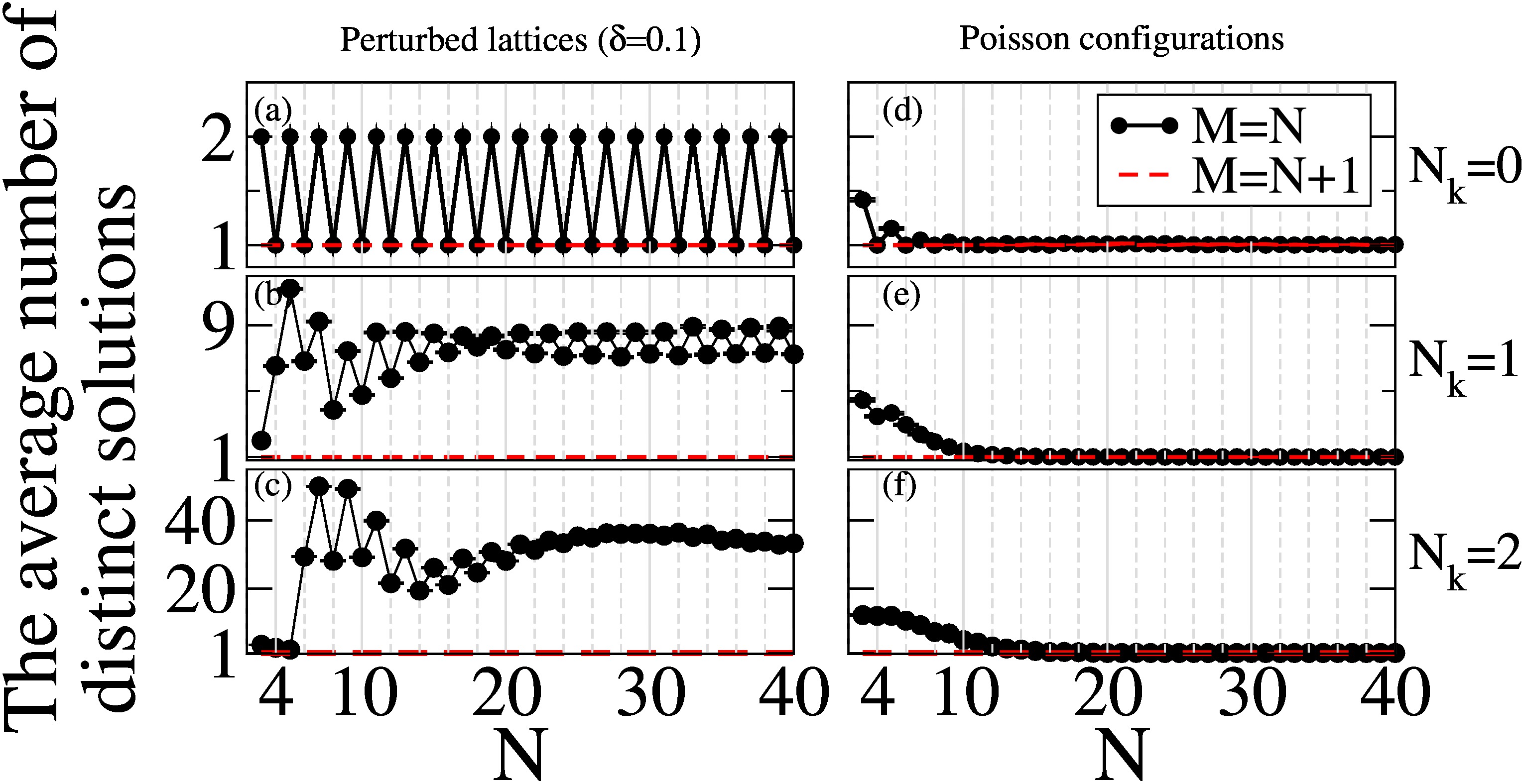}
\end{center}
\caption{
Numerical results for the average number of numerically distinct solutions per a target configuration of particle number $N$ with various values of $N_k$.
Using the real condition \eqref{cond:real} and BFGS+MINOP algorithms, we consider two types of target configurations: (a-c) perturbed lattices with $\delta=0.1$ and the tolerance $\epsilon_E = 10^{-20}$, and (d-f) Poisson configurations with $\epsilon_E = 10^{-25}$.
When $N_k=0$, both types of target configurations require $M=N$ constraints for an even-number $N$, and $M=N+1$ is the minimal for an odd-number $N$:  The minimal number of $M$ is $2\ceil{N/2}$.
If $N_k>0$, for both types of target configurations, the minimal number of constraints becomes $M=N+1$.
\label{fig:numSols_large systems}}
\end{figure}

Here, we numerically investigate the properties of the inversion procedure from collective coordinates, such as proper values of the tolerances $\epsilon_E$ and $\epsilon_X$.
For this purpose, we obtain distributions of energy $\fn{\Phi}{\vect{r}^N;\vect{R}^N}$ for numerically distinct solutions, as we did in figure \ref{fig:energy distribution 3prt}.
Our results, shown in figures \ref{fig:energy distributions} and \ref{fig:energy distribution2}, demonstrate that the energy distributions sensitively depend on the number of skipped collective-coordinate constraints $N_k$ as well as target configurations and the particle number $N$.

At first, we consider the cases with $N_k = 0$ (figure \ref{fig:energy distributions}).
When there are even-number $N$ of particles, $M\geq N$ constraints can give unique solutions for both types of target configurations: perturbed lattices and Poisson point distribution configurations.
If $N$ is an odd number, however, $M=N$ constraints no longer ensure unique solutions.
When perturbed lattices are the target configurations (figure \ref{fig:energy distributions}(a-c)) and $M=N$ constraints are considered, the energy $\fn{\Phi}{\vect{r}^N;\vect{R}^N}$ always has two global minima, which correspond to the trivial solution ($\vect{r}^N =\vect{R}^N$) and a nontrivial one ($\vect{r}^N \neq \vect{R}^N$), respectively.
On the other hand, the energy $\fn{\Phi}{\vect{r}^N;\vect{R}^N}$ of a Poissonian target configuration (figure \ref{fig:energy distributions}(d-f)) mostly has a single minimum  that is identical to the target ($\vect{r}^N = \vect{R}^N$) but occasionally has more than two nontrivial solutions.
Given parameters $N_k=0$ and $M=N+1$, while when the target is a perturbed lattice the inversion procedure gives a unique solution, when the target is a Poisson configuration this procedure may give multiple solutions.
However, since the nontrivial solutions in the latter case have qualitatively different energy profiles from the trivial solution (see figure \ref{fig:energy distributions}(d-f)), the nontrivial solutions can be eliminated by lowering the tolerance $\epsilon_E$ to a proper level. Thus, when $N$ is an odd number, $M=N+1$ constraints are required for the unique determination.

When first few collective coordinates are skipped ($N_k >0$), there is no advantage of even-number particles, i.e., one cannot determine unique solutions with $M=N$ successive collective-coordinate constraints when $N$ is an even number.
Figure \ref{fig:energy distribution2} shows the histograms for energies of numerical solutions obtained in the inversion procedure with an odd-number particles and $N_k>0$.
In figure \ref{fig:energy distribution2}, we note that for $M=N$ constraints there can be more than one nontrivial solutions whose energy profiles are similar to that of the trivial solutions.
However, $M=N+1$ constraints allow us to find the trivial solutions without any nontrivial one.

In general, as the system size $N$ increases, both trivial and nontrivial solutions tend to have higher energies, i.e., larger numerical errors.
Moreover, for parameters $N_k=0$ and $M=N$, although for smaller systems the distribution of trivial and nontrivial solutions have tails in the low-energy regime  [figure \ref{fig:energy distributions} (a, d)], for larger systems the tails are shifted to the high-energy regime  [figure \ref{fig:energy distributions} (c, f)]; see also figure \ref{fig:energy distribution2} for cases with $N_k>0$.
This observation implies that it becomes less probable to obtain numerical solutions, whether they are trivial or not, as the particle number $N$ increases, or the energy tolerance $\epsilon_E$ is lowered.

\begin{figure}[h]
\begin{center}
\includegraphics[width = 0.8\textwidth]{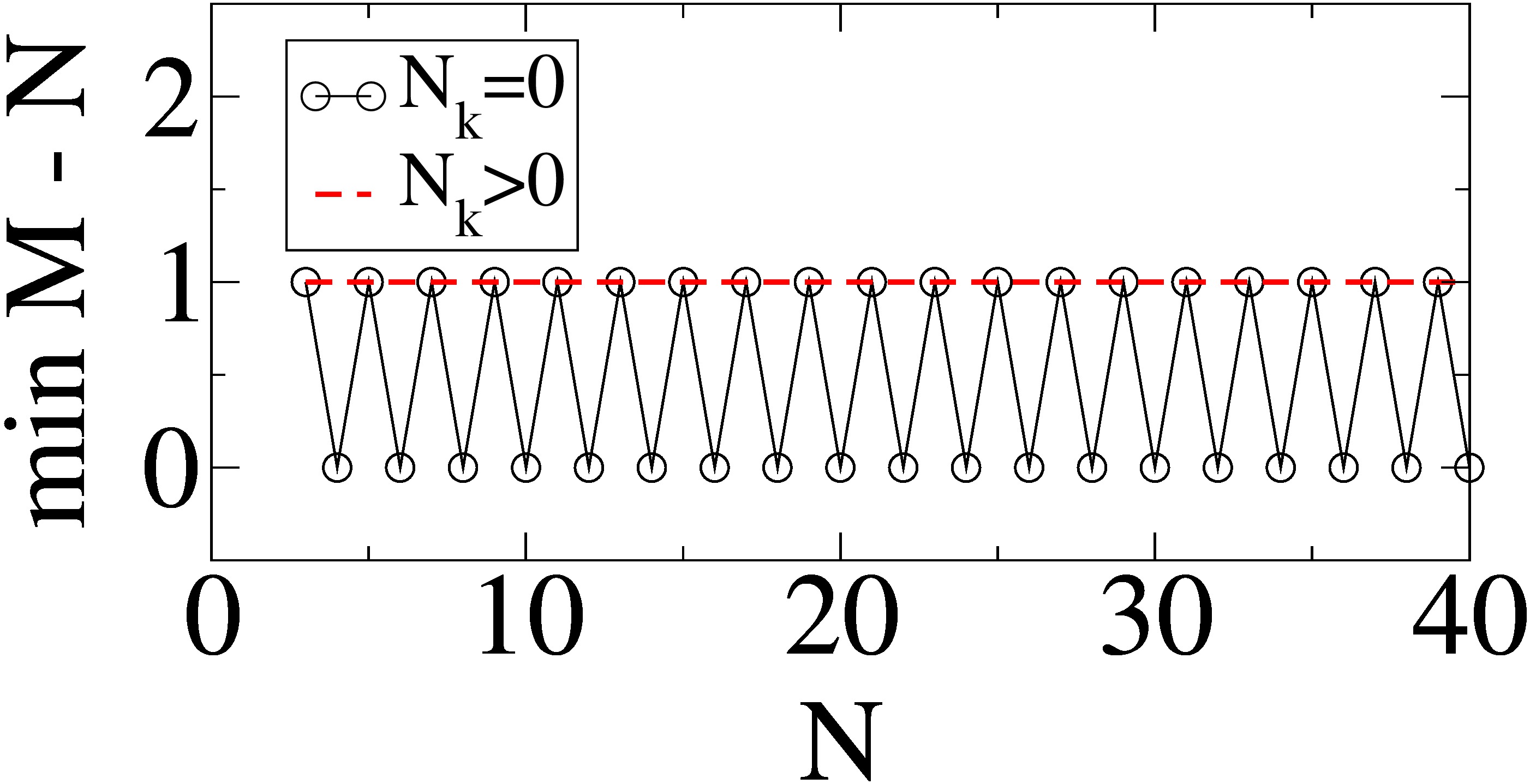}
\end{center}
\caption{The minimal number of successive collective-coordinate constraints $\min M$ as a function of particle number $N$ for various $N_k$. \label{fig:conclusion}}
\end{figure}

The average number of numerically distinct solutions, obtained in the inversion procedure, is shown in figure \ref{fig:numSols_large systems}.
This figure clearly demonstrates that  for Poissonian targets (figure \ref{fig:numSols_large systems}(d-f)) the two curves ($M=N$ and $N+1$) collapses into a single line as $N$ increases, and thus $\min M \to N$ as $N$ increases.
On the other hand, these two curves are separated for perturbed lattices (figure \ref{fig:numSols_large systems}(a-c)), and thus $\min M$ is determined by the cases where perturbed lattices are the target configurations.
Figure \ref{fig:conclusion} summarizes the results from analytic investigation into small systems (section \ref{sec:small systems}) and numerical studies on larger systems (section \ref{sec:N>3}).
One can uniquely determine particle coordinates from collective coordinates at the $\ceil{\frac{N}{2}}$ smallest wavenumbers, i.e., parameters of $N_k=0$ and $M = 2\ceil{\frac{N}{2}}$, by properly selecting $\epsilon_E$.
On the other hand, if $N_k>0$, one requires $M=N+1$ successive collective-coordinate constraints to uniquely determine particle coordinates.
Therefore, when both cases are considered, the minimal set of collective-coordinate constraints are collective coordinates at the $\ceil{\frac{N}{2}}$ smallest wavenumbers.

\section{Conclusions and Discussions}\label{sec:conclusions}

In this work, we have investigated the minimal set of collective-coordinate constraints as a function of the particle number $N$ to uniquely determine the progenitor particle coordinates in one dimension.
We also considered how the minimal collective-coordinate constraints depend on constraint types (the real \eqref{cond:real} and imaginary \eqref{cond:imag} conditions) and types of target configurations, i.e., perturbed lattices and Poisson point distribution configurations.
As shown in figure \ref{fig:conclusion}, the minimal set of constraints are collective coordinates at the $\ceil{\frac{N}{2}}$ smallest wavenumbers: It corresponds to the parameters of $N_k=0$ and $M=2\ceil{N/2}$.
In other words, the removed number of degrees of freedom in the solution space will vary with each collective-coordinate constraint, and  the real and the imaginary parts of a collective coordinate are not completely independent.

For this result to accommodate the pathological case, i.e., the integer lattice, one needs to regard all of its translations to be equivalent.
As we noted in section \ref{sec:N=2}, this is because translations of the integer lattices cannot be distinguished in terms of $\colT{k_m}$ for $m=1,\cdots, \ceil{N/2}$, since their collective coordinates are identically zero, except at the Bragg peaks, i.e., $k = 2\pi, 4\pi, \cdots$.
An additional constraint $\colT{k_N} \equiv \colT{2\pi}$ at the first Bragg peak is necessary to remove the translational degree of freedom.
However, we note that non-Bravais lattices are not pathological cases because their lattice constants are larger than one, and thus their first Bragg peaks should appear within the range of $\abs{k}\leq \pi$.

It is worthwhile to compare this conclusion with the result of Fan \textit{et al.} \cite{Fan1991}.
These authors proved that for a one-dimension system one needs its collective coordinates at the $\floor{\frac{N}{2}}$ smallest wavenumbers as well as the center of mass in order to determine all of its collective coordinates; see \ref{sec:Fan's result} for the detailed summary.
In the same context, our investigation shows that if the center of mass is unknown, one needs collective coordinates at the $\ceil{\frac{N}{2}}$ smallest wavenumbers.
Moreover, when there are an even-number of particles, the knowledge of the center of mass does not reduce the necessary information.

\begin{figure}[h]
\begin{center}
\includegraphics[width = 0.8\textwidth]{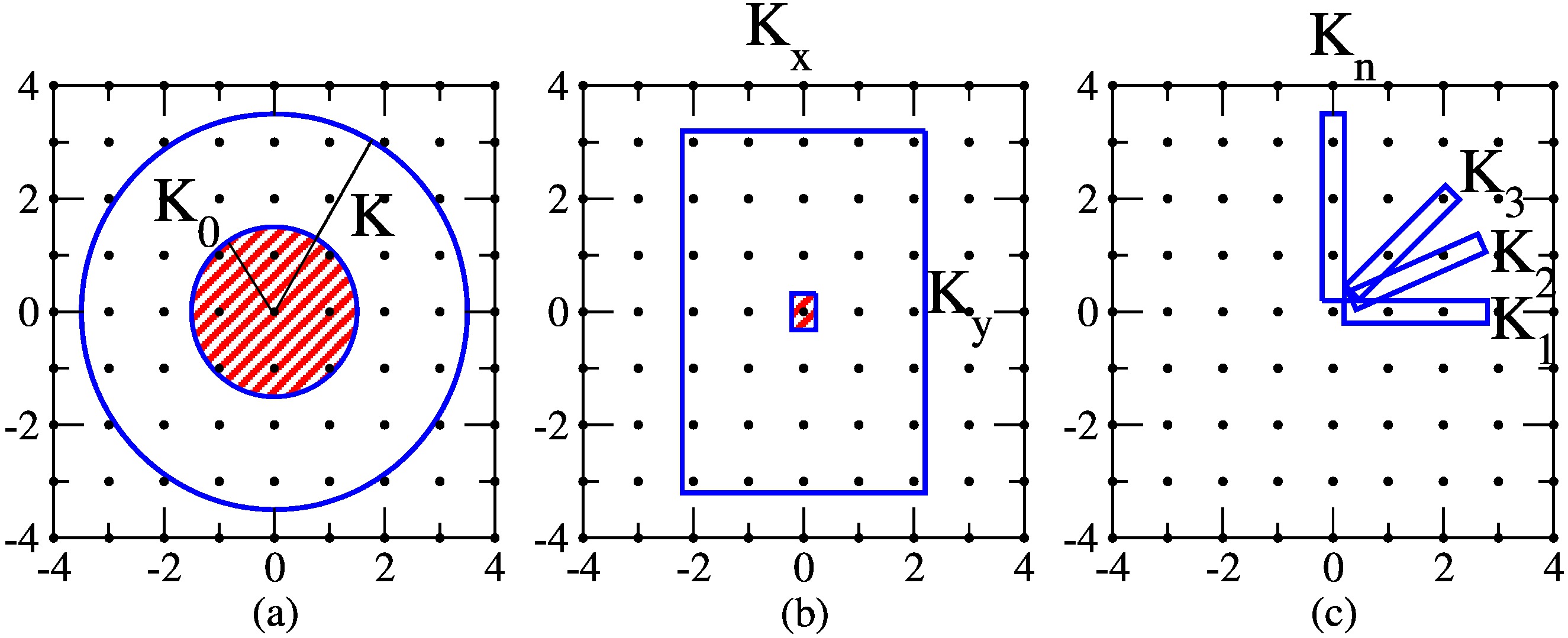}
\end{center}
\caption{Schematics of some possible ways to select collective-coordinate constraints in the two-dimensional Fourier space.
Collective coordinates are specified at wavevectors inside (a) an annular region of outer radius $K$ and inner radius $K_0$ (see figure \ref{fig:Constraints}), 
(b) a rectangular region of width $K_x$ and height $K_y$, and (c) $n$ mutually non-parallel strips which lengths are $K_i$, $i=1,\cdots, n$.
We note that the red-shaded region is excluded. \label{fig:inclusion regions}}
\end{figure}

While the present work focused on one-dimensional systems for simplicity, it is useful to discuss implications of our results for the inversion problem in higher-dimensional systems.
Unlike one-dimensional systems, higher-dimensional systems can have many different ways to select collective-coordinate constraints; see figure \ref{fig:inclusion regions}.
Here, consider the case (c) where selected wavevectors form $n$ nonparallel strips orienting toward the origin.
Based on our present results, if the $i$th strip has a slope $s_i = n_i/m_i$, where $n_i$ and $m_i$ are integers and coprime, and includes the smallest $\ceil{N/2}$ wavevectors, then one can uniquely determine values of the coordinates on a line, i.e., $m_i x_j + n_i y_j$ for $j=1,\cdots, N$.
Thus, by using two perpendicular strips that include a total of $2\ceil{N/2}$ collective-coordinate constraints, one can ``separately" determine the $x$ and $y$ coordinates of particle positions.
In order to determine the pairing between the $x$ and $y$ coordinates, one needs collective-coordinate constraints along additional strips in the Fourier space, as shown in figure \ref{fig:inclusion regions}(c).
Therefore, in this scheme at least $3\ceil{N/2}$ collective-coordinate constraints are required.

It is interesting to compare collective coordinates with Fourier components in discrete Fourier transform (DFT).
While a Fourier component $X_k$ in DFT is a linear function of a complex sequence $\{x_n\}_{n=0}^{\mathcal{N}-1}$, a collective coordinate $\col{k_m}$ is a nonlinear function of particle coordinates $\{R_j\}_{j=1}^N$.
In both cases, wavenumbers are restricted to be equally spaced due to the periodic boundary conditions in direct spaces.
On the contrary, the direct spaces are different in the two cases in that while the direct spaces in DFT are digitized into $\mathcal{N}$ pixels, those in collective coordinates are continuous.
%{\color{green}For unique inversions, $\mathcal{N}/2$ Fourier components or $\ceil{N/2}$ collective coordinates are required, respectively.}
If one discretizes the space of a point configuration with $\mathcal{N}$ pixels of width $\Delta x$, the configuration can be described by a real-valued sequence $\{x_n \}$, where $x_n$ represents the number of particles in the $n$th pixel.
Then, this conversion can be straightforwardly written as follows:
\begin{center}
\begin{tabular}{r c c c}
Particle coordinates: & $\{R_i\}_{i=1}^N \subset \R$ &$\Rightarrow$ &$\{x_n\}_{n=0}^{\mathcal{N}-1} \subset \N \cup \{0 \}$ \\ 
Collective coordinates: & $\col{k_m} = \sum_{i=1}^N \exp(-i k_m x_i	)$ & $\Rightarrow$ & $X_m = \sum_{j=0}^{\mathcal{N}-1} x_j \exp\left[-i\frac{2\pi  m}{\mathcal{N}\Delta x} (j\Delta x)\right] $.
\end{tabular}
\end{center}
Thus, the $m$th collective coordinate $\col{k_m}$ of a point configuration corresponds to the $m$th Fourier component $X_m$ of its digitized version. 
From this relationship, one can surmise that  the inverse DFT with the first $\mathcal{N}/2$ collective coordinates will give  a discretized point configuration with a position precision $\Delta x$.
%if  are performed via the inverse DFT, then the result will represent a discretized point configuration with a position precision $\Delta x$.
In other words, one needs around $10^{7}$ Fourier components to achieve $\Delta x \sim \order{10^{-7}}$, which is a typical error in our solution configurations.

In the present work, we focused on the search for the minimal set of constraints, rather than computational costs.
Our inversion procedure is intuitive and provides easy-to-estimate numerical errors in solutions (i.e., energy $\fn{\Phi}{\vect{r}^N;\vect{R}^N}$), but this method is inefficient for large systems.
For instance, as system size $N$ increases, the computation cost grows at least in the order of $N^2$.
Furthermore, since this method tends to have larger numerical errors in solution configurations as $N$ increases (see figures \ref{fig:energy distributions} and \ref{fig:energy distribution2}), it becomes more likely to fail to find any solution with a given value of the energy tolerance $\epsilon_E$.
The failure rate becomes especially much higher when a target is more complicated.
% because the tolerance $\epsilon_E$ should be sufficiently lowered to obtain a unique solution.
Therefore, for future studies, it would be important to develop more efficient procedures to invert collective coordinates into particle coordinates.

\section*{Acknowledgement}

This work was supported partially by the National Science Foundation under Grant No. CBET-1701843.

\appendix
\section{Approximate Solutions of Equations \eqref{eq:Cn_N=3} and \eqref{eq:Sn_N=3}}

For parameters $N_k=0$ and $M=3$, and the real condition \eqref{cond:real}, from \eqref{eq:Cn_N=3} and \eqref{eq:Sn_N=3}, one can find two solutions as follows:
\begin{eqnarray}
\fl \delta_1 &\approx \frac{\left(	-18\Imag{\colT{k_1}} \pm D\right)}{\pi(\Real{4\colT{k_1}-\colT{k_2}})\left(\Real{4\colT{k_1} -\colT{k_2}}-12	\right)} \nonumber \\
\fl\delta_2 &\approx -\frac{\Real{4\colT{k_1}-\colT{k_2}}}{12\sqrt{3}\pi} + \frac{6}{\Real{4\colT{k_1}-\colT{k_2}} -6}\left(\frac{\Imag{\colT{k_1}}}{2\pi}-\delta_1	\right)\label{eqs:sol_3prt_1}\\
\fl\delta_3 &\approx \frac{\Real{4\colT{k_1}-\colT{k_2}}}{12\sqrt{3}\pi} + \frac{6}{\Real{4\colT{k_1}-\colT{k_2}} -6}\left(\frac{\Imag{\colT{k_1}}}{2\pi}-\delta_1	\right), \nonumber
\end{eqnarray}
where the discriminant $D$ is written as
\begin{eqnarray}
\fl D \equiv & \frac{1}{12\sqrt{3}} \left(	\Real{4\colT{k_1} -\colT{k_2}}-6\right) \Big[ \left(	\left(	\Real{4\colT{k_1} -\colT{k_2}}-6\right)^2-36 \right) \nonumber \\
\fl &\times  \left(\Real{4\colT{k_1} -\colT{k_2}}{}^2 -36\Real{2\colT{k_1}+\colT{k_2}}	\right) +3888\Imag{\colT{k_1}}{}^2     \Big]^{1/2}
. \label{eq:D-term} 
\end{eqnarray}
Here, a trivial solution is obtained  from \eqref{eqs:sol_3prt_1} when a minus sign is taken in $\delta_1$.
Otherwise, \eqref{eqs:sol_3prt_1} become a nontrivial solution.

For parameters $N_k=1$ and $M=4$, \eqref{eq:Cn_N=3} and \eqref{eq:Sn_N=3} give a single solution{\color{red}:}
\begin{eqnarray}
\fl\delta_2 & = -\frac{\delta_1}{2} + \frac{\Imag{\colT{k_3}}}{12\pi} +\left[\sqrt{3} \pi {\delta_1}^2 + \frac{1}{4\sqrt{3}\pi} \left(\Real{\colT{k_2} + \frac{2}{9}\colT{k_3}}-\frac{2}{3} 	\right)\right] \label{eq:sol_Nk=1_M_4(2)}\\
\fl\delta_3 & = -\frac{\delta_1}{2} + \frac{\Imag{\colT{k_3}}}{12\pi} -\left[\sqrt{3} \pi {\delta_1}^2 + \frac{1}{4\sqrt{3}\pi} \left(\Real{\colT{k_2} + \frac{2}{9}\colT{k_3}}-\frac{2}{3} 	\right)\right], \label{eq:sol_Nk=1_M_4(3)}
\end{eqnarray}
where $\delta_1$ is determined by the following cubic equation:
\begin{eqnarray}
\fl {\delta_1}^3 &-\frac{\Imag{\colT{k_3}}}{6\pi}{\delta_1}^2  -\frac{1}{12\pi ^2} \left(\Real{\colT{k_2}+\frac{2}{9}\colT{k_3} }-\frac{11}{3}\right) \delta_1 \nonumber \\ 
\fl &+\frac{1}{72 \pi^3}\left[\Imag{\colT{k_3} }\left(\Real{\colT{k_2} +\frac{2}{9}\colT{k_3} }-\frac{5}{3}	\right)	-3 \Imag{\colT{k_2}}\right] =0\label{eq:sol_Nk=1_M_4(1)},
\end{eqnarray}
which has a single real root.

%In terms of positions of a target configuration $\Delta_i$ for $i=1,2,$ and $3$, solutions \eqref{eqs:sol_3prt_1} can be written as
%\begin{eqnarray}
%\delta_1 &  = -\Delta_1 + \frac{3\pi ({\Delta_2}^2-{\Delta_3}^2) +\sqrt{3}(-2\Delta_1+\Delta_2+\Delta_3)}{3\pi (\sqrt{3}\pi (\Delta_2 -\Delta_3) +2)(\Delta_2 -\Delta_3)} \\
%\delta_2 & = \Delta_2+\frac{-6 \pi  \Delta_1 \left(\Delta_2-\Delta_3\right)+\sqrt{3} \left(-2 \Delta_1+\Delta_2+\Delta_3\right)}{3 \pi  \left(\sqrt{3}\pi   \left(\Delta_2-\Delta_3\right)+2\right) \left(\Delta_2-\Delta_3\right)}\\
%\delta_3 & = \Delta_3+\frac{-6 \pi  \Delta_1 \left(\Delta_2-\Delta_3\right)+\sqrt{3} \left(-2 \Delta_1+\Delta_2+\Delta_3\right)}{3 \pi  \left(\sqrt{3}\pi   \left(\Delta_2-\Delta_3\right)+2\right) \left(\Delta_2-\Delta_3\right)},
%\end{eqnarray}
%and 
%\begin{equation}
%D \approx \frac{ -\left(\sqrt{3}\pi (\Delta_2-\Delta_3)+1	\right)\abs{2\Delta_1 \left(\sqrt{3}\pi (\Delta_2-\Delta_3)+1	\right)-(\Delta_2 +\Delta_3)}}{2\sqrt{3} \pi(\Delta_2-\Delta_3) (\sqrt{3}\pi (\Delta_2 -\Delta_3)+2)}.
%\end{equation}
%In other words, as long as not all of $\Delta_i$ are equal, there exists at most two solutions as we observed in numerical studies.

\section{The uniqueness of solutions for the inversion problem}\label{sec:Fan's result}

Using the generating function argument \cite{Fan1991}, one can prove that there is the unique configuration to satisfy $N$ prescribed collective coordinates.
Let us define a generating function as 
\begin{equation}
\fn{f}{z} \equiv \sum_{m=1}^\infty \frac{\col{k_m}}{m} z^m,
\end{equation}
which is well-defined for $\abs{z} < 1$ because $\abs{\col{k_m}}$ is bounded.
Using the definition \eqref{def:collective coordinates} and power series expansion of the log function [$\ln(1-z) = \sum_{n=1}^\infty z^n /n$ for $\abs{z}<1$], 
\begin{eqnarray}
\fn{f}{z} &= \sum_{n=1}^\infty \left( \sum_{j=1}^N {e^{-inx_j}}	\right)\frac{z^n}{n}  = \sum_{j=1}^N \sum_{n=1}^\infty \frac{(ze^{-ix_j})^n}{n} = \sum_{j=1}^N -\fn{\ln}{1-ze^{-ix_j}} \nonumber\\
& = -\ln \left[\prod_{j=1}^N (1-ze^{-ix_j})	\right]. \label{eq:gen_function1}
\end{eqnarray}
Since the term inside square brackets of logarithm is a polynomial of order $N$, $\exp{[f(z)]}$ also should be a polynomial of order $N$.
\begin{eqnarray}
\prod_{j=1}^N (1-ze^{-ix_j}) & = \fn{\exp}{-\fn{f}{z}}  = \mathbb{P}_N \fn{\exp}{-\fn{f}{z}} = \mathbb{P}_N \fn{\exp}{-\mathbb{P}_N \fn{f}{z}}\nonumber \\
&= \mathbb{P}_N \fn{\exp}{- \sum_{m=1}^N \frac{\col{k_m}}{m} z^m},\label{eq:generating_function2}
\end{eqnarray}
where $\mathbb{P}_N$ represents a projection to a degree $N$ polynomial of $z$.

By substituting \eqref{eq:generating_function2} into \eqref{eq:gen_function1} and doing further analysis, Fan, \textit{et al.} \cite{Fan1991} derived the following identity:
\begin{equation}\label{eq:generating function_Fan}
\fl\sum_{m=1}^N \frac{\col{k_m}}{m} z^m = -\ln\left[\mathbb{P}_{\floor{\frac{N}{2}}} \fn{\exp}{-\sum_{m=1}^{\floor{N/2}} \frac{\col{k_m}}{m} z^m} -\omega z^N \mathbb{P}_{-\floor{\frac{N}{2}}} \fn{\exp}{\sum_{m=1}^{\floor{N/2}} \frac{\col{-k_m}}{m} z^{-m} }  	\right],
\end{equation}
where $\omega \equiv \fn{\exp}{-i 2\pi \sum_{n=1}^N x_n}$, and $\floor{x}$ is the floor function of  $x$.
Since $\col{k_m} = \col{-k_m}^*$, if collective coordinates at the $\floor{\frac{N}{2}}$ smallest wavenumbers and the center of mass are known, in principle one can determine collective coordinates at other wavenumbers.
In other words, there is a unique point configuration that satisfy these conditions.

\section*{References}
\providecommand{\newblock}{}

\end{document}